\documentclass[aps, pra, twocolumn]{revtex4-1}
\usepackage[unicode=true,pdfusetitle, bookmarks=true,bookmarksnumbered=false,bookmarksopen=false, breaklinks=false,pdfborder={0 0 0},backref=false,colorlinks=false] {hyperref}
\hypersetup{ colorlinks,linkcolor=myurlcolor,citecolor=myurlcolor,urlcolor=myurlcolor}
\usepackage{graphics,graphicx, amsthm, amsmath, amssymb, times, braket, colortbl, color, bm, soul}
\usepackage[up]{subfigure}
\usepackage{cleveref}
\usepackage[outdir=./]{epstopdf}
\definecolor{myurlcolor}{rgb}{0,0,0.7}

\theoremstyle{plain}

\providecommand{\theoremname}{Theorem}

\newcommand*{\myproofname}{Proof}

\makeatother

\begin{document}

\title{Population inversion and entanglement in single and double glassy Jaynes-Cummings models}


\author{Ahana Ghoshal}
\affiliation{Harish-Chandra Research Institute, HBNI, Chhatnag Road, Jhunsi, Allahabad 211 019, India}

\author{Sreetama Das}
\altaffiliation[Present address: ]{Faculty of Physics, Arnold Sommerfeld Center
for Theoretical Physics, Ludwing Maximilian University Munich,
Theresienstr. 37, 80333 Munich, Germany}
\affiliation{Harish-Chandra Research Institute, HBNI, Chhatnag Road, Jhunsi, Allahabad 211 019, India}

\author{Aditi Sen(De)}
\affiliation{Harish-Chandra Research Institute, HBNI, Chhatnag Road, Jhunsi, Allahabad 211 019, India} 

\author{Ujjwal Sen}
\affiliation{Harish-Chandra Research Institute, HBNI, Chhatnag Road, Jhunsi, Allahabad 211 019, India} 

\begin{abstract}

We find that a suppression of the collapse and revival of population inversion occurs in response to insertion of Gaussian quenched disorder in atom-cavity interaction strength in the Jaynes-Cummings model. The character of suppression can be significantly different in the presence of non-Gaussian disorder, which we uncover by studying the cases when the disorder is uniform, discrete, and Cauchy-Lorentz. Interestingly, the quenched averaged atom-photon entanglement keeps displaying nontrivial oscillations even after the population inversion has been suppressed. Subsequently, we show that disorder in atom-cavity interactions helps to avoid sudden death of atom-atom entanglement in the double Jaynes-Cummings model. We identify the minimal disorder strengths required to eliminate the possibility of sudden death. We also investigate the response of entanglement sudden death in the disordered double Jaynes-Cummings model in the presence of atom-atom coupling.

\end{abstract}

\maketitle
\section{Introduction}
The Jaynes-Cummings (JC) model, developed by Jaynes and Cummings in 1963 \cite{JC}, is possibly the simplest model describing a two-level quantum system, interacting with a single mode of a quantized field. In this model, the field interacting with the two-level system was treated quantum mechanically, in contrast to the (semiclassical) methods where the field was 
treated classically. 
Evidence of oscillation of coherence of this model and decay of the oscillation amplitude, was noted by Cummings \cite{cummings}, for short time intervals. In \cite{Narozhny}, the authors demonstrated periodic collapses and succeeding revivals, at larger times, of the atomic population. They found an expression for a short-time ``collapse function", and observed
periodical long-time revivals and 
a monotonically decreasing envelope of the revivals. In \cite{Rempe}, the quantum collapse and revival conjectured in the JC model were verified experimentally. The time evolution for relatively short times of population inversion in the JC model has periodic collapse-revival nature \cite{Scully}; however, at later times, there appear fractional revivals, i.e., revivals that occur at a time which is a noninteger multiple of the time interval between revivals near the initial time and that has a smaller amplitude than that of the initial revivals \cite{Averbukh, Dooley}. The superstructure (revivals, fractional revivals, and super-revivals) of this model was also described later in 1993 \cite{Gora}. The nature of atom-photon entanglement in this model was studied in \cite{Tannor}. See also \cite{Knight}. 
Evolution of the atomic density matrix of a dissipative JC model was studied in \cite{Wonderen}. The authors of that
work also proposed a limit for which the 
matrix converges to a state of maximum von Neumann entropy.
 The entanglement dynamics of a double Jaynes-Cummings model was studied in \cite{Yu, Masood, Chan, Wang, Vieira}.  The phenomenon of entanglement sudden death (ESD) was observed in the double JC model 
 in \cite{Almeida, Eberly}. For further studies, see
 \cite{Chen, Pandit, Zhang, Shen, Sainz, Hu, Liao, Tanas}.\par
In realistic scenarios, the fabrication of a perfect cavity with the atom sitting at a locked position inside the cavity is quite challenging. There are copious possibilities of fluctuations in tuning the components of the set-up, which pushes the actual experiment away from the ideal scenario, typically described in the literature. This difficulty gives rise to disorder in the parameters, contributing to alteration in the dynamics of the atom-cavity system, which in turn changes the nature of dynamics of various physical properties associated to the system. Our aim in this paper is to look into such effects of disorder on the dynamics of JC models. More precisely, we find the response to disparate strains of quenched disorder in the atom-cavity interaction parameter on population inversion and atom-photon entanglement in a single atom-cavity system described by the JC model, and the atom-atom entanglement in a double JC model.\par
In Sec. \ref{sect-ek}, we briefly describe the JC model and the phenomenon of collapse and revival of population inversion. In Sec. \ref{Que}, the concept of quenched disorder and quenched averaging is discussed in the context of the Jaynes-Cummings model with a quenched disordered atom-cavity coupling constant. The coupling constant is assumed to be respectively affected by four paradigmatic forms of disorder distributions, viz., Gaussian, uniform, discrete and Cauchy-Lorentz. The concepts of median and semi-interquartile range, required for analyzing the Cauchy-Lorentz disordered parameter, are also reviewed. Sections \ref{sec_3} and \ref{sec_4} consider the response of population inversion to the different types of disorder. Section \ref{sec_5} briefly recapitulates the dynamics of atom-photon entanglement in the clean JC model. In Sec. \ref{sec_6}, we present the results about the response of atom-photon entanglement to a disordered interaction in the JC model. In Sec. \ref{sec_7}, we consider the double JC model, focusing on the behavior in time of atom-atom entanglement. We separately consider the cases which exhibit entanglement sudden death and the ones which do not. In each case, we find the response of atom-atom entanglement to quenched disordered atom-cavity interactions. In particular, in Sec. \ref{sub_2}, we locate the regions in the system parameter spaces, formed by the strengths of the disorders in the interactions and the initial-state entanglement, that allow entanglement sudden death even after quenched averaging and that preclude the same. In Sec. \ref{atom-atom}, we investigate the effects of the Gaussian quenched disordered atom-photon coupling term on the sudden death of atom-atom entanglement in the double JC model in the presence of additional atom-atom coupling terms in the Hamiltonian. A conclusion is presented in Sec. \ref{9}.   
\section{The Jaynes-Cummings Model}
\label{sect-ek}
\begin{figure}
\includegraphics[width=9cm,height=7cm]{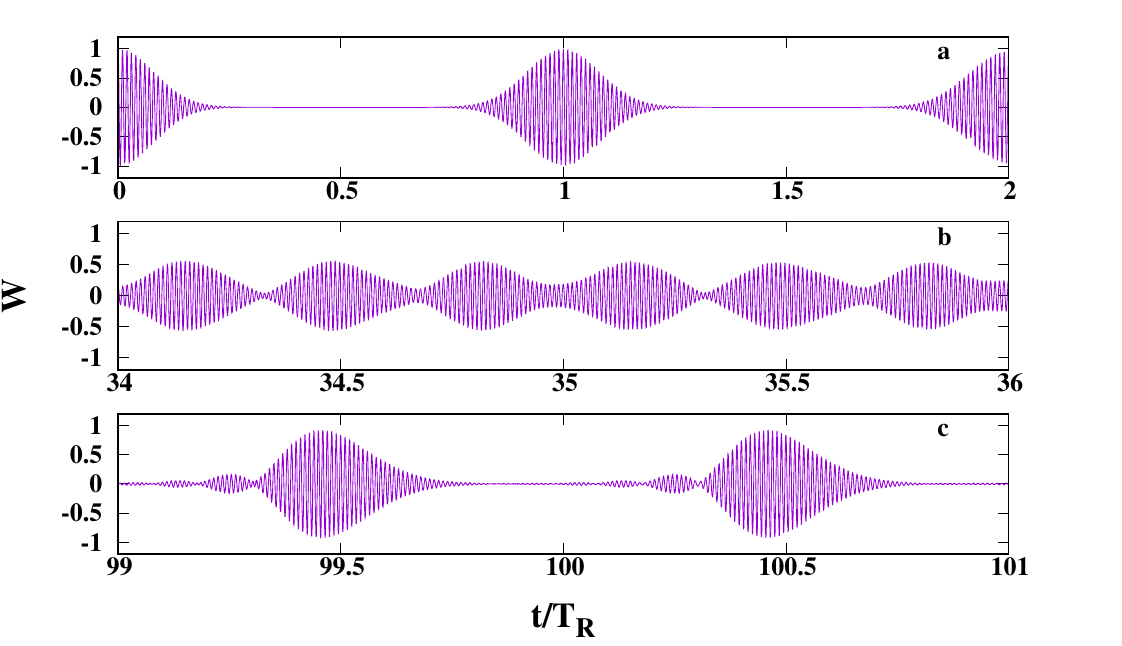}
 \caption{Various types of revivals in time evolution of the JC model. The population inversion is plotted on the vertical axes against time on the horizontal ones, for $\bar{n}=50$ and $\Delta n=2$. The revivals in the early times in $(a)$ are quite different from those in the later ones. So, there are revivals of period $\frac{T_{R}}{3}$ in $(b)$ (fractional revivals), and super-revivals for $99T_{R}<t< 101 T_{R}$ in $(c)$. All axes represent dimensionless parameters. See text for references where this figure was previously plotted and analyzed.}
\label{fig-ek} 
 \end{figure}

The interaction Hamiltonian of a single mode of a quantized field of frequency $\nu$ with a single two-level atom in the JC model is given by  
\begin{equation}
    H_{I}=\hbar g(|1\rangle\langle 0|a+|0\rangle\langle 1|a^{\dagger} ),
\end{equation}
where $|0\rangle$ and $|1\rangle$ are the ground and excited states of the two-level atom; $a$ and $a^{\dagger}$ are, respectively, the annihilation and creation operators of the photon field mode; and $g$ is the coupling strength between the cavity (realizing the mode) and the atom. The total Hamiltonian of a single Jaynes-Cummings set-up can be represented as
\begin{equation}
H = \frac{\hbar\omega}{2}\sigma_{z} + \hbar\nu a^{\dagger}a + H_{I},
\end{equation}
where $\hbar\omega$ is the energy difference between the atomic levels and $\sigma_{z}=|0\rangle\langle 0|-|1\rangle\langle 1|$.\\
The initial states of the field and atom are, respectively, chosen as
\begin{equation} \nonumber
|\psi(0)\rangle_{field}=\sum_{n=0}^{\infty}C_{n}|n\rangle
\end{equation}
and
\begin{equation} \nonumber
|\psi(0)\rangle_{atom}=\alpha|0\rangle+\beta|1\rangle.
\end{equation} 
The states $|n\rangle, n=0,1,2,....,$ of the mode are the photon number states. $C_n, \alpha,$ and $\beta$ are complex constants.\\
The initial atom-cavity joint state is 
\begin{equation}
|\psi(0)\rangle=\sum_{n=0}^{\infty}C_{n}(\alpha |0,n\rangle + \beta |1,n\rangle),
\end{equation}
where $ |0,n\rangle $ and $ |1,n\rangle $ are the states of the total system, having $ n $ photons in the field and the atom being in ground and excited states, respectively.
The eigenvectors evolve as \cite{Shen}
\begin{eqnarray}
&& |\psi_{1,n}(t)\rangle = e^{\frac{-iHt}{\hbar}}|1,n\rangle \nonumber \\
&& =\cos(gt\sqrt{n+1})|1,n\rangle-i \sin(gt\sqrt{n+1})|0,n+1\rangle
\label{2}
\end{eqnarray}
and
\begin{eqnarray}
&& |\psi_{0,n}(t)\rangle =e^{\frac{-iHt}{\hbar}}|0,n\rangle \nonumber \\
&& =\cos(gt\sqrt{n})|0,n\rangle-i \sin(gt\sqrt{n})|1,n-1\rangle,
\label{3}
\end{eqnarray} 
so that the wave function at some time $t$ \cite{Averbukh} reads as
\begin{eqnarray}
\nonumber
&& |\psi(t)\rangle= \\ \nonumber
&& \sum_{n=0}^{\infty}\big\lbrace[\beta C_{n}\cos(g\sqrt{n+1}t)-i\alpha C_{n+1}\sin(g\sqrt{n+1}t)]|1\rangle \\ \nonumber
&& +[-i\beta C_{n-1}\sin(g\sqrt{n}t) +\alpha C_{n}\cos(g\sqrt{n}t)]|0\rangle\big\rbrace |n\rangle
\label{4}
\end{eqnarray}
with the detuning parameter $\Delta=\omega-\nu$ being set to zero. Now, assuming that the atom is initially in its ground state i.e., $\alpha=1$ and $\beta=0$, the population inversion \cite{Gora} of this system is
\begin{equation}
     W ( t ) = \sum_{n=0}^{\infty} |C_{n}|^{2} \cos(2 g t \sqrt{n}).
\end{equation}
In the above expression, $|C_{n}|^{2}$ stands for the initial photon distribution. Here we consider a sub-Poissonian statistics, i.e., $\Delta n\ll\bar{n}$, with $\bar{n}$ and $\Delta n$ being the mean and standard deviation of photon distribution. As in \cite{Averbukh}, we choose the initial state of the field as a Gaussian distribution of $C_{n}$, viz., $C_{n}$ being real, and
\begin{equation}
C_{n}^{2}= \frac{1}{\sqrt{2 \pi} \Delta n} \exp [- \frac{(n-\bar{n})^{2}}{2 \Delta n^{2}}].
\label{C_n}
\end{equation}
The time evolution of the population inversion for an initially large average number of photons gives rise to the collapse and revival phenomenon. The revival period $T_{R}$ can be estimated as the time when the $\bar{n}^{th}$ and $(\bar{n}+1)^{th}$ components are in the same phase. The expression of $T_{R}$ for large $\bar{n}$ is \cite{Gora} 
\begin{equation}
     T_{R}= \frac{2 \pi \sqrt{\bar{n}}}{g}.
\label{5}     
\end{equation}
With the increase of time, fractional revivals and super-revivals can also occur \cite{Averbukh}. See Fig. {\ref{fig-ek}}.

 \section{The Jaynes-Cummings model with a quenched disordered coupling}
\label{Que} 

The presence of different types of randomness or disorder in the parameters describing the quantum system, or inhomogeneity in the medium of the system, can have a multitude of consequences, like the breakdown of periodicity in that medium. This in turn can result in localization behavior of the wave function, in certain regions of the medium. This localization, in the context of cooperative phenomena was first suggested by Anderson in \cite{anderson}, where diffusive waves were shown to vanish in the presence of a disordered medium. Introducing disorder can lead to an altered nature in behavior of different other physical quantities of the system. The disorder may be either inherent in the physical realization of the quantum system or artificially incorporated in the system. The relevant value of a physical quantity of a disordered system is obtained by an averaging over a large number of realizations of the disorder. Depending on the characteristic time-scales of the system and the disorder, there are two paradigmatic methods of this averaging - ``annealing'' and ``quenching''. We will discuss the second one, i.e., quenched disorder and the corresponding averaging, in the next subsection. In the rest of the paper, we consider different distributions of quenched disorder in the system parameters of the JC model, and study the response to it on characteristics like population inversion and entanglement of the system.\par
The investigation of effects of disorder in the Jaynes-Cummings model can hardly be overemphasized. It is often challenging to fabricate a perfect cavity with a uniform field inside, and a two-level atom with a fixed energy difference between its levels. In reality, it is more probable to have some disorder in the tuning of the parameters, due to the nonachievement of ideal experimental situations. Previous work on systems akin to the JC model in the presence of disorder includes \cite{Quach,Kulaitis,mascarenhas}. For investigations in the JC model with fluctuating coupling constants, see, e.g., \cite{Puri, Quang, Joshi2, Joshi, Lawande, Lawande1, Joshi1, Kayhan}.\par
The aim in this paper is to find the response, of population inversion and atom-photon entanglement in the JC model, and atom-atom entanglement in a double JC model, to quenched disorder in the atom-photon coupling(s). We will consider four types of quenched disordered couplings, three of which are continuous while one is discrete. Among the continuous ones, two have finite mean and standard deviation, while for the third, the quantities are undefined.
\subsubsection*{\textbf{Quenched disorder}}
  A system parameter is said to be quenched disordered when the equilibration time of the disorder in the system is much larger than the typical observation time that is being considered. This means that these disordered parameters, for a particular realization of the disorder, virtually do not change during the time of observation. They may change after a long time, but that range of time is not in the domain of our interest. Such systems are often referred to as ``glassy" \cite{Parisi}, and, likewise, we term a Jaynes-Cummings model with this type of disorder, as a glassy Jaynes-Cummings model.
\subsubsection*{\textbf{Models of disorder}}
 We consider an insertion of disorder in the atom-cavity coupling strength $ g $ in the interaction Hamiltonian $ H_{I} $. This is realistically justifiable because the fluctuation of position of the atom inside the cavity, if any, can give rise to a fluctuating $ g $. The interaction Hamiltonian with disorder is written as
  \begin{equation}
  \tilde{H_{I}}  =\hbar g (1+\delta) ( |1\rangle \langle 0|a+ |0\rangle \langle 1|a^{\dagger} ),
  \label{equ_9}
\end{equation}
where the disorder is modeled by $ \delta $, which we have taken to have different continuous and discrete distributions. Specifically, we have assumed the following types of the disorders.\\
$\bullet$ \textbf{Gaussian quenched disorder:} In this case, $\delta$ is chosen to be from a Gaussian distribution with zero mean and standard deviation, $s$, so that the corresponding probability density function is given by
\begin{equation}
P(\delta=\delta)= \frac{1}{s\sqrt{2\pi}}e^{-\frac{1}{2}(\frac{\delta}{s})^2}, \qquad -\infty\textless\delta\textless\infty. 
\end{equation}
We will often refer to the dispersion of a distribution of a disordered system parameter, as quantified by the standard deviation or the ``semi-interquartile range" (see below for a definition) of the distribution, to gauge (measure) the ``strength" of the disorder introduced.\\
$\bullet$ \textbf{Uniform quenched disorder:} In this instance, $\delta$ is distributed as
\begin{eqnarray}
 P(\delta=\delta)&=& 1/s   \qquad when\: -\frac{s}{2}\leq\delta\leq \frac{s}{2},\\ \nonumber
 &=& 0         \qquad otherwise.
\end{eqnarray}
$\bullet$ \textbf{Discrete quenched disorder:} In this case, $\delta$ is distributed as
  \begin{eqnarray}
 P(\delta=\delta)&=& \frac{1}{2}  \qquad when\:  \delta= \pm\frac{s}{2},\\ \nonumber
 &=& 0         \qquad otherwise.
\end{eqnarray}
Note that, unlike the other cases, we have considered here a discrete probability distribution.\\
$\bullet$ \textbf{Cauchy-Lorentz quenched disorder:} This disorder is differently distributed as compared to the previous three types of disorder, because the mean does not exist for this distribution. In this instance, $\delta$ is distributed as
   \begin{equation}
P(\delta=\delta)= \frac{s}{\pi}\frac{1}{\delta^2+s^2}, \qquad -\infty<\delta<\infty. 
\end{equation}
It is a continuous probability distribution, the mean of which, i.e., $\int_{-\infty}^{\infty}\delta P(\delta=\delta)d\delta$, does not exist.
\subsubsection*{\textbf{Median}}
The mean of a probability distribution is a very important measure of central tendency. However, in instances when it does not exist and in some other cases, it is fruitful to consider the median  \cite{Gupta}. The median $M$ of a continuous probability distribution $P(\delta)$ is its middlemost value, and is given by
\begin{equation*}
\int_{-\infty}^{M} P(\delta=\delta)d\delta=\frac{1}{2}.
\end{equation*}
One can similarly identify the first and third quartiles, respectively, as
\begin{equation*}
\int_{-\infty}^{Q_1} P(\delta=\delta)d\delta=\frac{1}{4} \qquad \textrm{and} \qquad \int_{-\infty}^{Q_3} P(\delta=\delta)d\delta=\frac{3}{4},
\end{equation*}
with the median being the second quartile. The semi-interquartile range, $\frac{1}{2}(Q_3-Q_1)$, may be asked to play the role of the standard deviation, being another measure of dispersion of the probability distribution. For a discrete probability distribution, $P(A=a_i)=p_i$, of a random variable $A$, the median can be defined as $\frac{1}{2}(\tilde{a_r}+\tilde{a}_{r+1})$, if $\sum\limits_{i\leq r}\tilde{P_i}<\frac{1}{2}<\sum\limits_{i>r}\tilde{P_i}$, and as $\tilde{a_r}$, if $\sum\limits_{i\leq r}\tilde{P_i}=\frac{1}{2}$, where $\lbrace\tilde{a_i}\rbrace$ is an ordered set, arranged in ascending or descending order and is equal to $\lbrace a_i\rbrace$ as a set. And, $P(A=\tilde{a_i})=\tilde{p_i}$. 
\subsubsection*{\textbf{Quenched averaging}}
A physically relevant value of a system characteristic of a disordered physical system is obtained by a suitable averaging over the disorder. If the disorder parameters are quenched disordered, the averaging has to be performed only after all other relevant operations have already been carried out. In particular, for finding the quenched averaged atom-photon entanglement of the system described by the JC Hamiltonian $\tilde{H_I}$ of Eq. (\ref{equ_9}), we first evaluate the entanglement $E_{\delta}(t)$, of the relevant quantum state $\psi_\delta(t)$ for an arbitrary but fixed time, \(t\). The quenched averaged entanglement is then given by
\begin{equation*}
\int_{-\infty}^{\infty} E_{\delta}(t) P(\delta=\delta)d\delta,
\end{equation*}
where the integral is to be replaced by a sum for discrete probability distributions. If such an integral or sum cannot be handled analytically, we take recourse to a numerical method. Typically, we will then Haar uniformly generate $N$ instances of the disorder $\delta$, and if they are referred to as $\delta_i$,
the quenched averaged entanglement will be
\begin{equation*}
\frac{1}{N}\sum_{i=1}^{N} E_{\delta_i}(t),
\end{equation*}
with the $N$ being sufficiently large that convergence, till a certain precision, with respect to $N$ has been reached. This avenue for finding the quenched averaged quantity, however, pre-assumes that the integrals and sums converge to finite values, which may not be guaranteed, in general, and especially for probability distributions without a finite mean. Therefore, if $P(\delta)$ is a Cauchy-Lorentz distribution, we will consider the median of the set $\lbrace E_{\delta_i}(t)\rbrace_{i=1}^N$ as the quenched averaged entanglement, where $N$ is again chosen to be sufficiently high so that convergence, till a certain precision, with respect to $N$ has been reached. 
\section{Localization of population inversion for Gaussian quenched disorder}
\label{sec_3}
A Gaussian distributed disorder has a very prominent effect, as compared to other types of disorder, on the dynamics of population inversion of the JC model. Suppose we choose $\delta$ as a random variable from the Gaussian distribution with zero mean and standard deviation $ s $. So, the population inversion for the Gaussian disorder after quenched averaging, is
 \begin{equation*}    W_{G} ( t ) = \sum_{n=0}^{\infty} C_{n}^{2} \int_{-\infty}^{\infty} \cos(2 g (1+\delta) t \sqrt{n}) \frac{e^{-\frac{\delta ^{2}}{2 s ^{2}}}}{s \sqrt{2 \pi}} d \delta
\end{equation*}
\begin{equation*}
 =\sum_{n=0}^{\infty} C_{n}^{2} \cos(2 g t \sqrt{n}) e^{-2 n s^{2} g^{2} t^{2}}.
\end{equation*}
Using Eq. (\ref{5}), we get
\begin{equation}
W_{G}(t)=\sum_{n=0}^{\infty} C_{n}^{2} \cos(2 g t \sqrt{n}) e^{-8 \pi^{2} n \bar{n} s^{2}(\frac{t}{T_{R}})^{2}}.
\label{7}
\end{equation}

\begin{figure*}
\includegraphics[width=18cm,height=8cm]{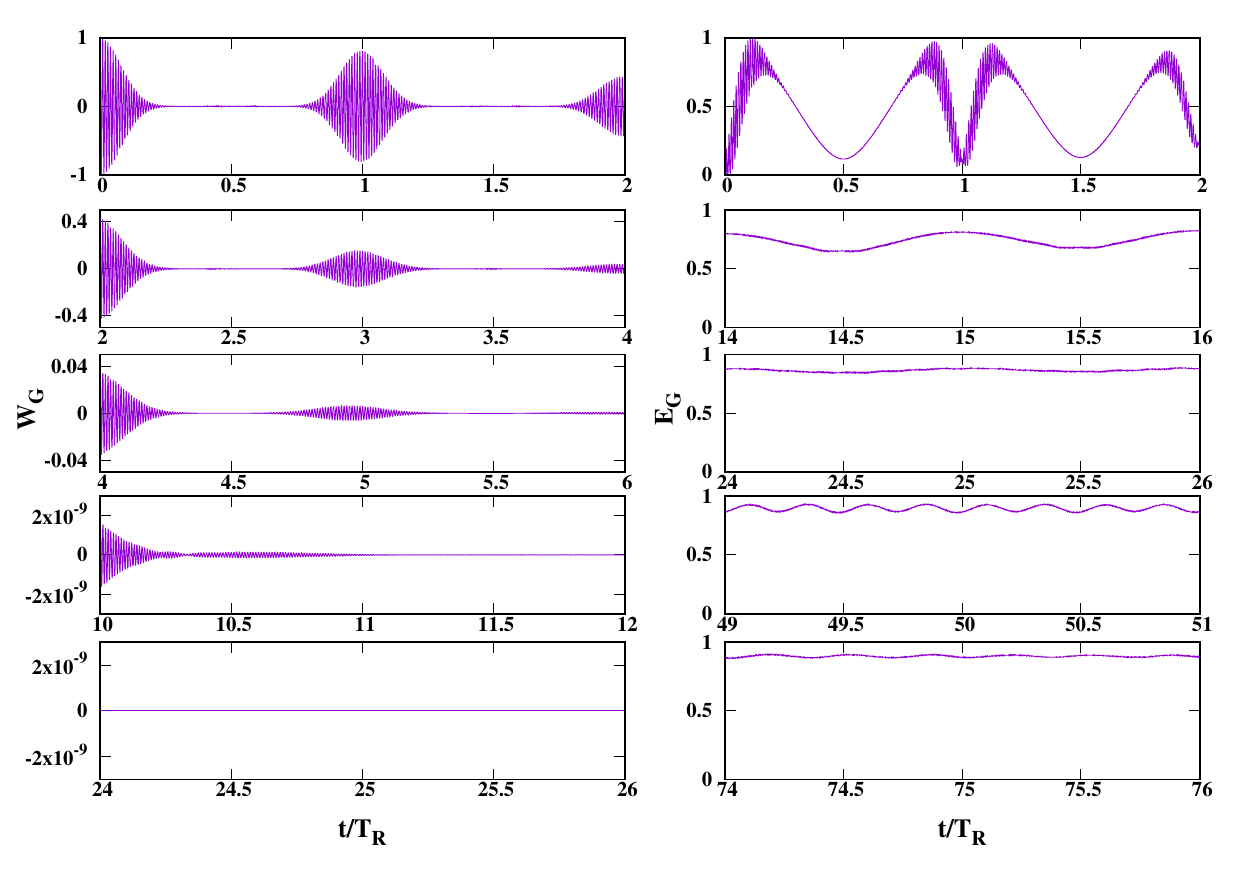}
\caption{Left panel: Population inversion gets strongly localized in time in response to Gaussian quenched disorder in atom-photon interaction of the JC model. Quenched averaged population inversion is plotted on the vertical axes against time on the horizontal ones. The Gaussian disorder has mean zero and standard deviation$=0.001$. All axes represent dimensionless quantities. Just like in Fig. \ref{fig-ek}, $\bar{n}=50$ and $\Delta n=2$, for the initial photon distribution. The notation for the quenched average is $W_G$ here.\\
Right panel: Response of quenched averaged atom-photon entanglement to quenched Gaussian disorder in the time evolved state of the JC model. All considerations are the same as in Fig. \ref{fig-dui} (left panel), except that the vertical axes are of entanglement, as quantified by local von Neumann entropy, and measured in ebits.}
\label{fig-dui}  
\end{figure*}
As seen in Fig. \ref{fig-dui} (left panel), we obtain a very sharp decay in the collapse-revival phenomenon of population inversion in the presence of Gaussian quenched disorder. After a moderately long time, the revivals completely disappear. In the figure, the decaying nature has been depicted for a very small standard deviation, namely, 0.001,  of the Gaussian distribution of $\delta$ (the mean being kept as vanishing), because for larger standard deviation the decay is very strong and we could not find any appreciable revivals. The threshold standard deviation for which there is no appreciable revival is approximately 0.005. Note that \(\delta\) is a dimensionless quantity. The collapse-revival phenomenon, therefore, is rather strongly ``localized in time" in response to Gaussian disorder in the atom-photon coupling. 
\section{Response of Population Inversion to non-Gaussian disorder}
\label{sec_4}
\noindent $\bullet$ \textbf{Uniform quenched disorder:} When $\delta$ is chosen randomly from a uniform distribution in the interval $[-s/2,s/2]$, the quenched averaged population inversion is
\begin{multline}
  W_{U} ( t ) = \sum_{n=0}^{\infty} C_{n}^{2} \int_{-\frac{s}{2}}^{\frac{s}{2}} \cos(2 g (1+\delta) t \sqrt{n}) \frac{1}{s} d\delta  \\
=\sum_{n=0}^{\infty} C_{n}^{2}\frac{1}{sgt\sqrt{n}}\cos(2gt\sqrt{n})\sin(sgt\sqrt{n}) \\
=\sum_{n=0}^{\infty} C_{n}^{2}\frac{1}{2\pi s \sqrt{n\bar{n}}\frac{t}{T_{R}}}\cos(4\pi\sqrt{n\bar{n}}\frac{t}{T_{R}})\sin(2 \pi s \sqrt{n\bar{n}}\frac{t}{T_{R}}).
\label{8}
\end{multline}
%
\begin{figure*}
\includegraphics[width=18cm,height=8cm]{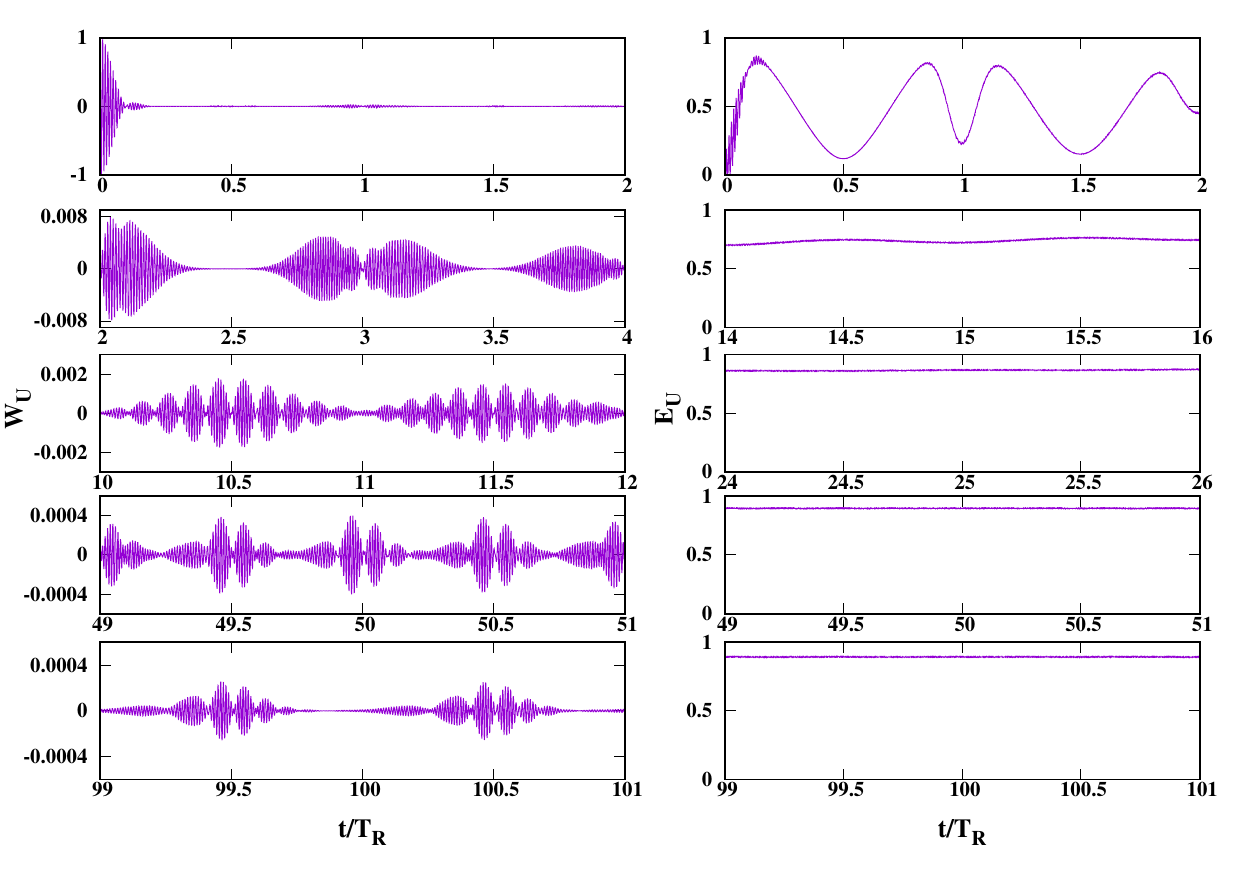}
\caption{Left panel:Population inversion gets localized for uniform quenched disorder, but not as strongly as for Gaussian disorder. All considerations are the same as in Fig. \ref{fig-dui} (left panel), except that the disorder is uniform, and although it still has mean$=0$ it has standard deviation$=0.1$. The notation for the quenched average is $W_U$ here.\\
Right panel:Effects of quenched uniform disorder on entanglement in time evolution within the JC model. All considerations remain the same as in Fig. \ref{fig-tin} (left panel), except that the vertical axes are now of entanglement and measured in ebits.}
\label{fig-tin} 
 \end{figure*}
%
%
Comparing Eq. (\ref{7}) with Eq. (\ref{8}), we see that while the suppression of the revivals in case of Gaussian quenched disorder was exponential in time it is only an inverse power suppression for uniform quenched disorder. See Fig. \ref{fig-tin} (left panel) for a depiction.\par
\noindent $\bullet$ \textbf{Discrete quenched disorder:} Here $\delta$ is chosen to take the values $-\frac{s}{2}$ and $\frac{s}{2}$ randomly but with the same probability. So, the quenched averaged population inversion becomes
\begin{multline}
 W_{D} ( t )= \sum_{n=0}^{\infty} C_{n}^{2} \frac{1}{2}[\cos (2g\sqrt{n}(1-\frac{s}{2})t)+ \cos (2g\sqrt{n}(1+\frac{s}{2})t)] \\
=\sum_{n=0}^{\infty} C_{n}^{2}\cos (2g\sqrt{n}t)\cos (sg\sqrt{n}t) \\
=\sum_{n=0}^{\infty} C_{n}^{2}\cos(4\pi\sqrt{n\bar{n}}\frac{t}{T_{R}})\cos(2 \pi s \sqrt{n\bar{n}}\frac{t}{T_{R}}).
\end{multline}
%
\begin{figure*}
\includegraphics[width=18cm,height=8cm]{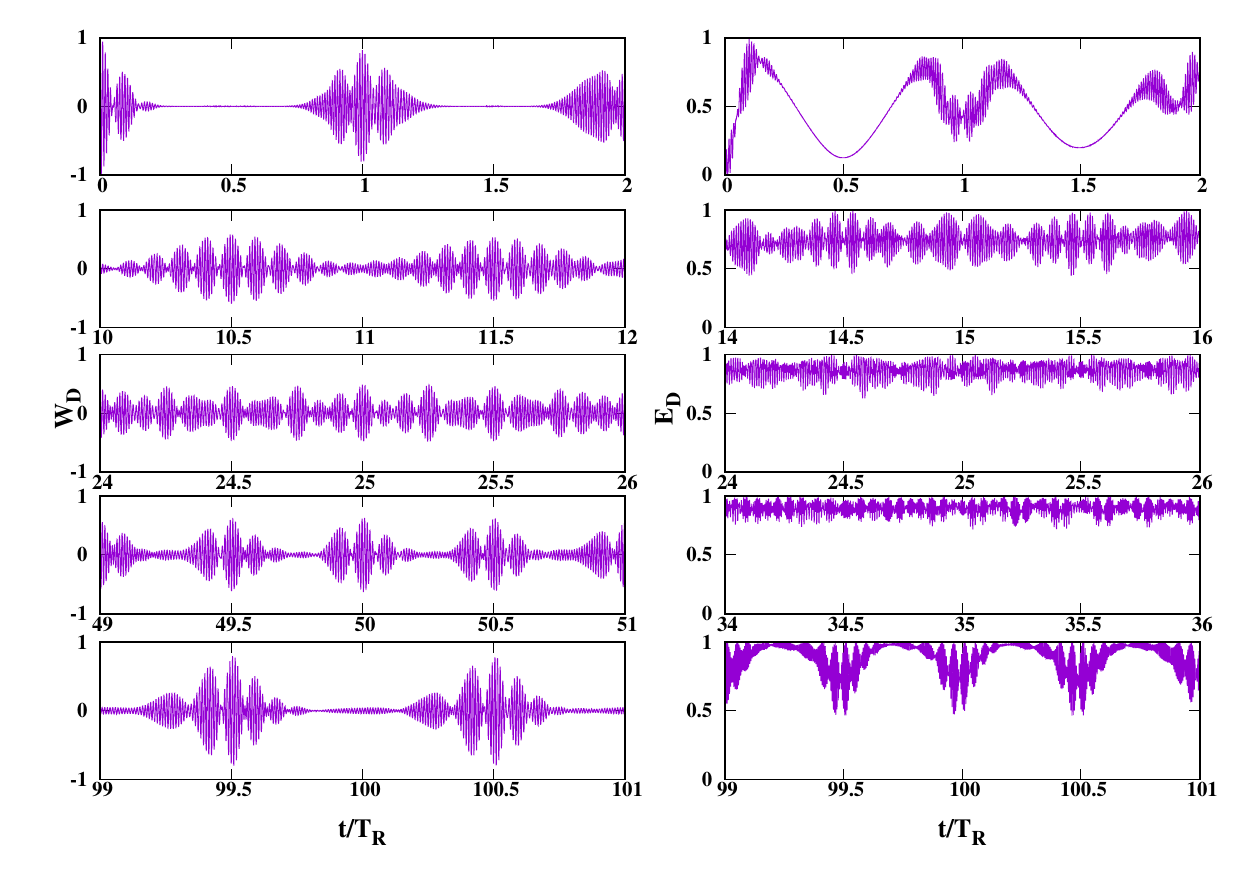}
\caption{Left panel: Population inversion retains much of its features even in the presence of disorder that is discrete in nature. All considerations are the same as in Fig. \ref{fig-tin} (left panel), except that the disorder is from the discrete distribution, as mentioned in the text. The notation for the quenched average is $W_D$ here.\\
Right panel: Effects of a quenched discrete disorder in coupling of the JC model on entanglement of the time evolved state. All considerations remain the same as in Fig. \ref{fig-char} (left panel), except that the vertical axes are now of entanglement, being measured in ebits.}
\label{fig-char}
\end{figure*}
%
%
%
From {Fig. \ref{fig-char}} (left panel), 
we can see that the amplitudes of the revivals are again suppressed, but the amount of suppression is very little, in comparison to those for Gaussian and uniform disorders. The original nature of the collapse and revival phenomenon is altered, but unlike the Gaussian and uniform disorders the fractional revivals and also the super-revivals are present in this case. Hence, it is plausible that \emph{discrete} quenched disorder does not leave a strong effect on the behavior of population inversion in a JC model.\par
\noindent $\bullet$ \textbf{Cauchy-Lorentz quenched disorder:} If $\delta$ is chosen from a Cauchy-Lorentz quenched disorder with vanishing median and semi-interquartile range $s$, we calculate the quenched averaged population inversion by finding the median of the distribution of population inversions for different realizations of the disorder. See Sec. \ref{Que} for the method of numerically estimating the median.
%
\begin{figure*}
\includegraphics[width=18cm,height=5cm]{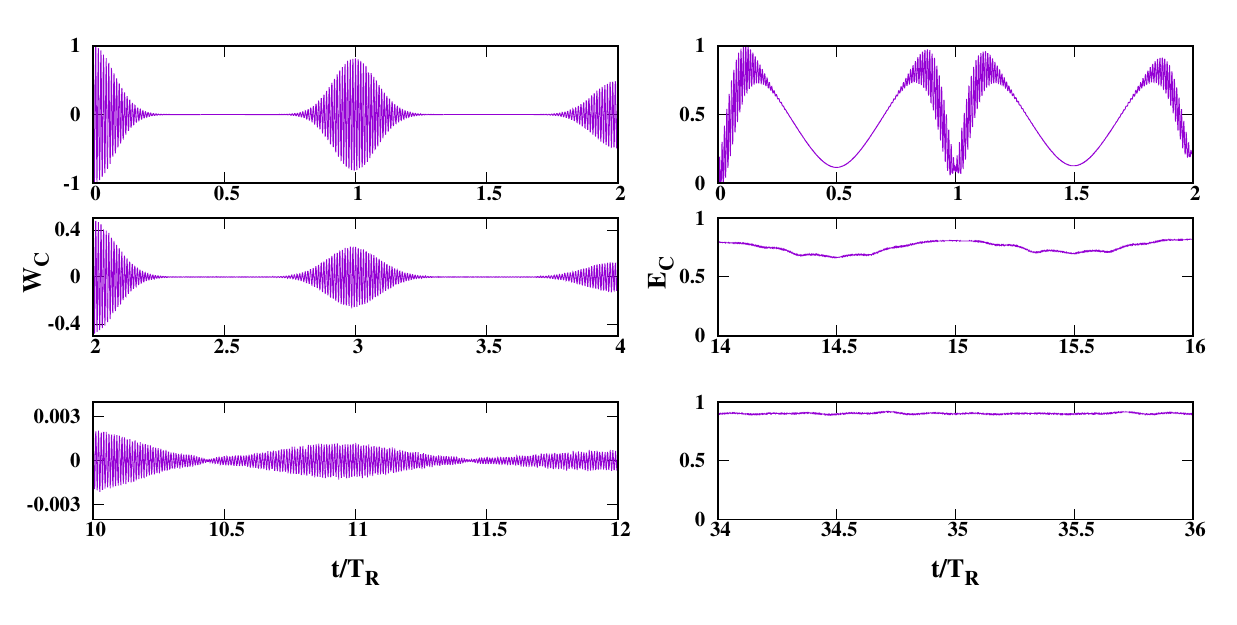}
\caption{Left panel: Median-based quenched averaged population inversion gets localized in time for quenched Cauchy-Lorentz disordered coupling in the JC model. All considerations are the same as in Fig. \ref{fig-dui} (left panel), except that the disorder distribution is Cauchy-Lorentz with its semi-interquartile range  $s=0.001$, and that the quenched averaging is the median-based one. The notation for the quenched averaged population inversion is $W_C$ here.\\
Right panel: Quenched Cauchy-Lorentz disorder in the JC model and its effect on entanglement in the time evolved state. All aspects remain the same as in Fig. \ref{fig-cau} (left panel), except that the vertical axes here are of entanglement and measured in ebits.}
\label{fig-cau}  
\end{figure*}
%
%
 %
Fig. \ref{fig-cau} (left panel) exhibits the nature of this median-based quenched averaged population inversion in the presence of quenched Cauchy-Lorentz disorder. This nature is quite similar to that for the quenched Gaussian disorder case.\par 
Among all the impurities considered here, we find that discrete quenched disorder provides the highest robustness to the phenomenon of collapse and revival of population inversion.
\section{Atom-photon entanglement in the Jaynes-Cummings model}
\label{sec_5}
If the population inversion of the atom-photon pure quantum state is $\pm 1$, the atomic state is in a pure state, precluding any atom-photon entanglement.
This holds irrespective of whether disorder is present in a system parameter, which would then necessitate disorder averaging. 
 The population inversion is seldom extremal, being instead much more often at zero. A vanishing population inversion in the atomic energy eigenbasis may imply a significant amount of the same in a basis complementary to the atomic energy basis, thereby again indicating zero entanglement. But it may also imply near-maximal atom-photon entanglement if all such population inversions are insignificant. In other words, studying the population inversion may not conclusively infer the complete information about atom-photon entanglement.\par
In this and the succeeding sections, we study the behavior of atom-photon entanglement with time in the JC model. The cases of the disordered couplings are dealt with in the succeeding section, while the ordered case is briefly described in this. The entanglement, being of a pure two-party quantum state at all times, can be measured by using the von Neumann entropy of any of the local density matrices. See \cite{Bennett} in this regard. We will see that the atom-atom entanglement, considered later in this paper and being for a mixed state in the time evolution of a double JC model, has to be measured differently.\par
Let us assume that initially the atom is in the ground state. Using Eq. (\ref{2}), we get the wave function of atom-photon system at time $t$, as
\begin{eqnarray}
&& |\psi(t)\rangle =\nonumber\\\nonumber&&\sum_{n=0}^{+\infty}[C_{n}\cos(g\sqrt{n}t)|0\rangle |n\rangle -iC_{n+1}\sin(g\sqrt{n+1}t)|1\rangle |n\rangle ].
\end{eqnarray}
After tracing out the field part, the reduced density matrix of the atomic subsystem is
\begin{equation}
        \rho=
  \left[ {\begin{array}{cc}
        a  & ib \\
       -ib & 1-a 
        \end{array} } \right],
\end{equation}
where
\begin{equation*}
a=\sum_{n=0}^{\infty} C_{n}^{2} \cos^{2}(g\sqrt{n}t),
\end{equation*}
\begin{equation*}
b=\sum_{n=0}^{\infty}C_{n}C_{n+1}\cos(g\sqrt{n}t)\sin(g\sqrt{n+1}t),
\end{equation*}
with $C_n$ being given by Eq. (\ref{C_n}).
The entanglement of the time evolved state is given by the von Neumann entropy of one of the reduced density matrices, i.e.,
\begin{equation}
E(t)=-\lambda_{1}\log_{2}\lambda_{1}-\lambda_{2}\log_{2}\lambda_{2},
\end{equation}
where $\lambda_{1}$ and $\lambda_{2}$ are the eigenvalues of $\rho$, given by
\begin{equation*}
\lambda_{1/2}=\frac{1}{2}(1\mp\sqrt{1+4b^{2}-4a+4a^{2}}).
\end{equation*}
\begin{figure}
\includegraphics[width=9cm,height=8cm]{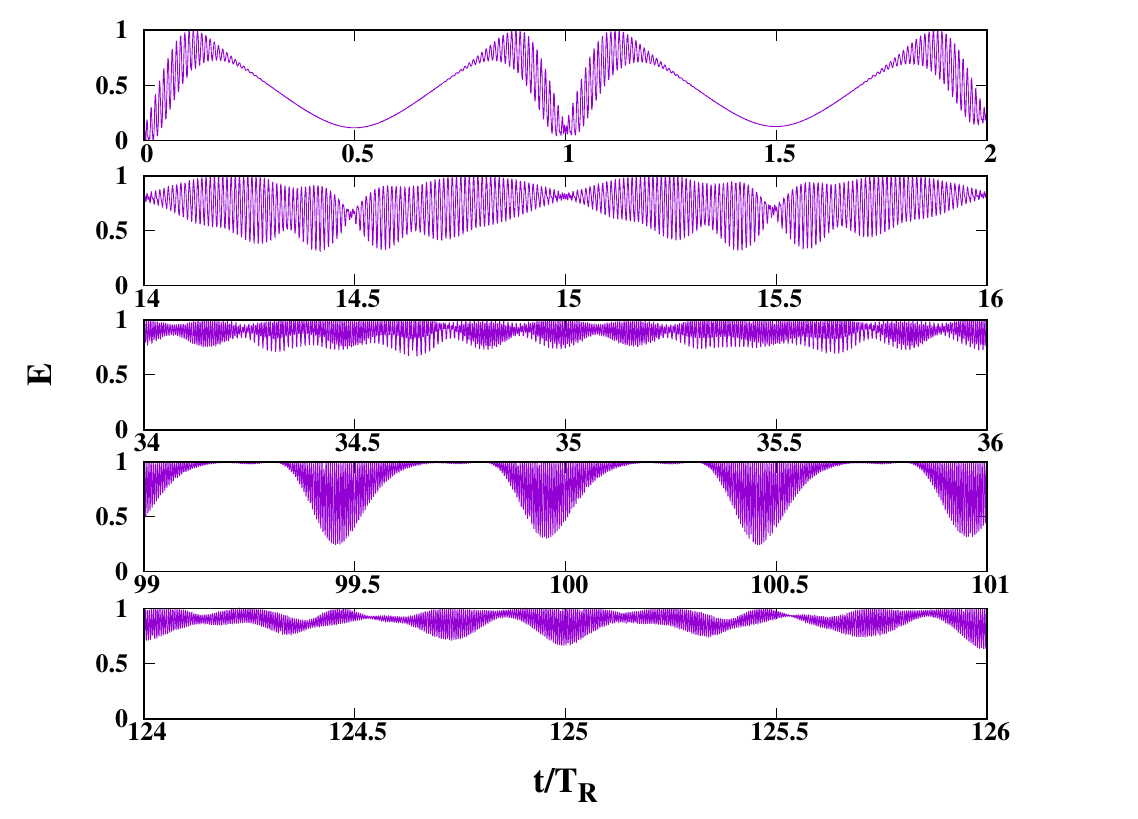} 
\caption{Time evolution of atom-photon entanglement in the JC model. The vertical axes are for entanglement and they are measured in ebits. The horizontal axes are dimensionless. All other considerations are the same as in Fig. \ref{fig-ek}.}
\label{ahana} 
\end{figure}
We plot the time dynamics of atom-photon entanglement in Fig. \ref{ahana}. We can see that initially the entanglement increases, in an oscillatory way, from 0 to 1, but with further increase of time the atom-photon system drifts far from the maximally entangled situation. After that, the entanglement again goes closer to 1 and this behavior repeats periodically. The behavior changes at even further times, when the entanglement may remain significantly close to maximal for long durations.\par
In the succeeding section, we will look at the behavior of entanglement in the time evolved state in the presence of different types of disorder.
\section{Response of atom-photon entanglement to disorder}
\label{sec_6}
Just as for the ordered case, it is almost never possible to infer the behavior of quenched averaged atom-photon entanglement from that of quenched averaged population inversion, 
obtained
in the disordered case. We study here the time dynamics of quenched averaged atom-photon entanglement for the same time evolution and same models of disorder for which we had examined population inversion. Since the disorder is assumed to be quenched, we average over the disorder \emph{after} evaluating the entanglements for given realizations of disorder.\par
\noindent $\bullet$ \textbf{Gaussian quenched disorder:} For a choice of disorder being randomly selected from a Gaussian distribution with zero mean and standard deviation $=0.001$, the short-time behavior of quenched averaged entanglement is significantly close to that for the clean case. 
Compare the top diagrams of Figs. \ref{ahana} and \ref{fig-dui} (right panel). This is rather close to the response of population inversion to Gaussian disordered couplings, at short times. See the top diagrams of Figs. \ref{fig-ek} and \ref{fig-dui} (left panel). For longer times, both population inversion and entanglement become relatively featureless in the disordered case, while the behaviors in the ordered case of both are rich in features. However, while the longer time quenched averaged population inversion is almost zero, the same of entanglement is almost maximal: while the latter implies the former, the reverse implication is not true.\par
We wish to make the following note about the choice of the value of standard deviation of the disorder. 
For the case of population inversion, we had set it to the relatively small value of 0.001, because higher values lead to complete suppression of the collapse-revival phenomenon of disorder-averaged population inversion. The situation is somewhat different for the disorder-averaged atom-photon entanglement. Higher values of the standard deviation keep the profile of the disorder-averaged entanglement to approximately the same as that for 0.001, with an interesting distinction. For standard deviation \(= 0.001\), there is a  strongly oscillating behavior of the profile of disorder-averaged entanglement within an oscillating envelope at near-initial times.
See the top diagram of the right panel of Fig. \ref{fig-dui}. For higher standard deviations, this strongly oscillating behavior  is suppressed, and occurs for very small initial times only.
\par
\noindent $\bullet$ \textbf{Uniform quenched disorder:} The effect of this disorder is quite similar to that of the Gaussian disorder, but for a higher value of the standard deviation.
Compare the right panels of Figs. \ref{fig-dui}  and \ref{fig-tin}. This is in contrast to the relative behaviors of population inversion in response to Gaussian and uniform disorders: quenched averaged population inversion had many more features for uniform disorder in comparison to that for Gaussian disorder. Compare the left panels of Figs. \ref{fig-dui} and \ref{fig-tin}. The standard deviation 
of the disorder has been chosen to be 0.1 for the diagrams in the panels of Fig. \ref{fig-tin}. 
The features of the envelope of the disorder-averaged entanglement remain approximately unaltered for other values of the standard deviation of the disordered interaction. However, for higher values of the standard deviation, the strong oscillations within the envelope for near-initial times are suppressed. 
\par
\noindent $\bullet$ \textbf{Discrete quenched disorder:} {Fig. \ref{fig-char}} (right panel) shows the nature of atom-photon entanglement in the presence of the discrete quenched disorder. In contrast to the previous cases of disorder, the discrete disorder retains many more features of the clean case, especially for longer times.\par 
\noindent $\bullet$ \textbf{Cauchy-Lorentz quenched disorder:} Fig. \ref{fig-cau} (right panel) shows the effect of this disorder. It is very similar to that due to Gaussian disorder, just as for population inversion.
\section{Entanglement in time evolution of quenched disordered Double Jaynes-Cummings model}
\label{sec_7}
In this section, we want to investigate the behavior of atom-atom entanglement in a double Jaynes-Cummings model. In a double Jaynes-Cummings model, there are two noninteracting atoms, each inside a cavity, and the cavities are isolated from each other. We will study the effect on the time evolution of an initially entangled state, between the two atoms, due to the presence of disorder in the atom-cavity interaction strengths. 
A note on the initial entanglement is in order here. The atoms are noninteracting for \(t\geq 0\), and therefore such an interaction cannot create any entanglement. Therefore, the initial entanglement is to be created by a 
mechanism that is independent of the Hamiltonian effective for \(t \geq 0\).\par
The double JC Hamiltonian is given by
\begin{eqnarray}
\nonumber
&& \mathcal{H}=\frac{\hbar\omega}{2} \sigma_{z}^{A}+\hbar(G_{A}\sigma_{+}^{A}a+G_{A}^{*}\sigma_{-}^{A}a^{\dagger})+\hbar \nu a^{\dagger}a \\ 
&& +\frac{\hbar\omega}{2} \sigma_{z}^{B} +\hbar (G_{B}\sigma_{+}^{B}b + G_{B}^{*}\sigma_{-}^{B}b^{\dagger}) + \hbar\nu b^{\dagger}b
\label{H}
\end{eqnarray}
where $\omega$ is the natural transition frequency between the excited state $|1\rangle$ and the ground state $|0\rangle$ of both the atoms, and $a^{\dagger}(a)$ and $b^{\dagger}(b)$ are the creation (annihilation) operators of the two single-mode fields with natural angular frequency $\nu$. $\sigma_+=|1\rangle\langle 0|$ and $\sigma_-=|0\rangle\langle 1|$, while $\sigma_{z}$ is the Pauli-$z$ operator. The superscripts $A$ and $B$ on them refer to the two atoms. $G_{A}$($G_B$) is the coupling strength between atom $A$ ($B$) and its cavity. The behavior of entanglement in the time dynamics of an initially entangled state when the dynamics is governed by the double JC Hamiltonian has been studied before. In particular the $G_{A}=G_{B}$ case was studied in \cite{Vieira, Eberly, Chen, Pandit, Zhang, Shen} and the $G_{A}\neq G_{B}$ case was studied in \cite{Shen, Sainz, Hu}. Here, we assume $\omega=\nu$ for simplicity, but keep $G_{A}\neq G_{B}$, in general, and study the response of atom-atom entanglement in the time evolved state to quenched disordered atom-cavity couplings.\par
In the ordered case, an interesting phenomenon uncovered a few years back in the time evolved state of the double JC model is ``entanglement sudden death" -- entanglement vanishes with a noncontinuous derivative with respect to time, and remains zero for a finite range of time for certain combinations of parameters in the initial state and the Hamiltonian. Below, we separately and briefly review the cases of the absence and presence of sudden death of entanglement in the clean cases, and after each case we correspondingly consider the response of the general properties of entanglement and of entanglement sudden death to quenched disorder in the coupling strengths of atom-cavity interactions.
\subsection{When sudden death of entanglement is not present in clean Hamiltonian}
\label{sub_1}
\subsubsection{Review of the clean case}
While the cases for which there is sudden death of entanglement in the double JC model are more appealing and have justly received more attention, there are also certain families of initial states for which the same model does not exhibit the phenomenon \cite{Eberly, Chen, Zhang, Shen, Sainz}. We briefly recapitulate the corresponding results.\par
Consider a partially entangled atomic pure state which is in the span of the two Bell states, $|\psi_{\pm}\rangle$, and is given by
\begin{equation}
|\psi_{atom}\rangle=\cos \alpha |1_{A},0_{B}\rangle+\sin \alpha |0_{A},1_{B}\rangle, 
\end{equation}
where $ \qquad|\psi_{\pm}\rangle=\frac{1}{\sqrt{2}}(|01\rangle\pm |10\rangle)$.\\
We assume that both the cavities are prepared initially in the vacuum states, $|0_{a}\rangle$ and $|0_{b}\rangle$.
So, the initial state for the total system is
\begin{equation}
|\psi_{0}\rangle=(\cos \alpha |1_{A},0_{B}\rangle+\sin \alpha |0_{A},1_{B}\rangle)\otimes |0_a0_b\rangle,
\label{si_0}
\end{equation}
where the suffixes $a$ and $b$ indicate states of the cavities interacting with atoms $A$ and $B$, respectively.
The evolved state of the double atom-cavity system can be written as
\begin{equation*}
|\psi_{t}\rangle=e^{-\frac{i\mathcal{H}t}{\hbar}}[(\cos \alpha |1_{A},0_{B}\rangle+\sin \alpha |0_{B},1_{A}\rangle)\otimes |0_a0_b\rangle]
\end{equation*}
Using Eqs.(\ref{2}) and (\ref{3}), we get
\begin{eqnarray}
\nonumber
|\psi_{t}\rangle=&&[\cos\alpha \lbrace \cos(G_{A}t)|1_{A},0_{a}\rangle\\\nonumber
&&-i \sin(G_{A}t)|0_{A},1_{a}\rangle \rbrace |0_{B},0_{b}\rangle+ \sin \alpha |0_{A},0_{a}\rangle\\
&&\times\lbrace \cos(G_{B}t)|1_{B},0_{b}\rangle -i \sin(G_{B}t)|0_{B},1_{b}\rangle \rbrace]
\end{eqnarray}
After tracing out the cavity parts, we have the density matrix for the two-atom system as
\begin{equation}
  \rho_{AB}(t)=
  \left[ {\begin{array}{cccc}
   0 & 0 & 0 & 0 \\
   0 & a & p & 0 \\
   0 & p^{*} & b & 0 \\
   0 & 0 & 0 & 1-a-b \\
  \end{array} } \right],
\end{equation}
where
\begin{eqnarray*}
a&=&\cos^{2} \alpha |\cos(G_{A}t)|^{2}\\
b&=&\sin^{2} \alpha |\cos(G_{B}t)|^{2}\\
p&=&\cos \alpha\sin \alpha \cos (G_A^{*}t)\cos(G_Bt).
\end{eqnarray*}
We now wish to evaluate the entanglement of the two-atom state. An information-theoretically meaningful entanglement measure is the entanglement of formation, which reduces to the local von Neumann entropy for pure bipartite states {\cite{Smolin}}. For two-qubit systems, the entanglement of formation is related to the ``concurrence" via a monotonically increasing function \cite{Wootters}, and, therefore, we use concurrence to measure the entanglement of $\rho_{AB}$, the state of the two atoms. The concurrence of a two-qubit state $\rho$ is given by
\begin{equation}
C=\max \lbrace 0,\sqrt{\lambda_1}-\sqrt{\lambda_2}-\sqrt{\lambda_3}-\sqrt{\lambda_4} \rbrace
\end{equation} 
where $\lambda_1$, $\lambda_2$, $\lambda_3$, and $\lambda_4$ are the eigenvalues, arranged in decreasing order, of the matrix $\rho \tilde{\rho}$, where
\begin{equation*}
\tilde{\rho}=(\sigma_{y}\otimes \sigma_{y})\rho^{*}(\sigma_{y}\otimes \sigma_{y}),
\end{equation*}
with $\sigma_y$ being the Pauli-$y$ matrix. Consequently, the concurrence of the two atoms is
\begin{equation}
C(t)=|\sin 2 \alpha \cos(G_{A}t)\cos(G_{B}t)|.
\label{banya}
\end{equation}
\begin{figure}
\includegraphics[width=9cm,height=6.5cm]{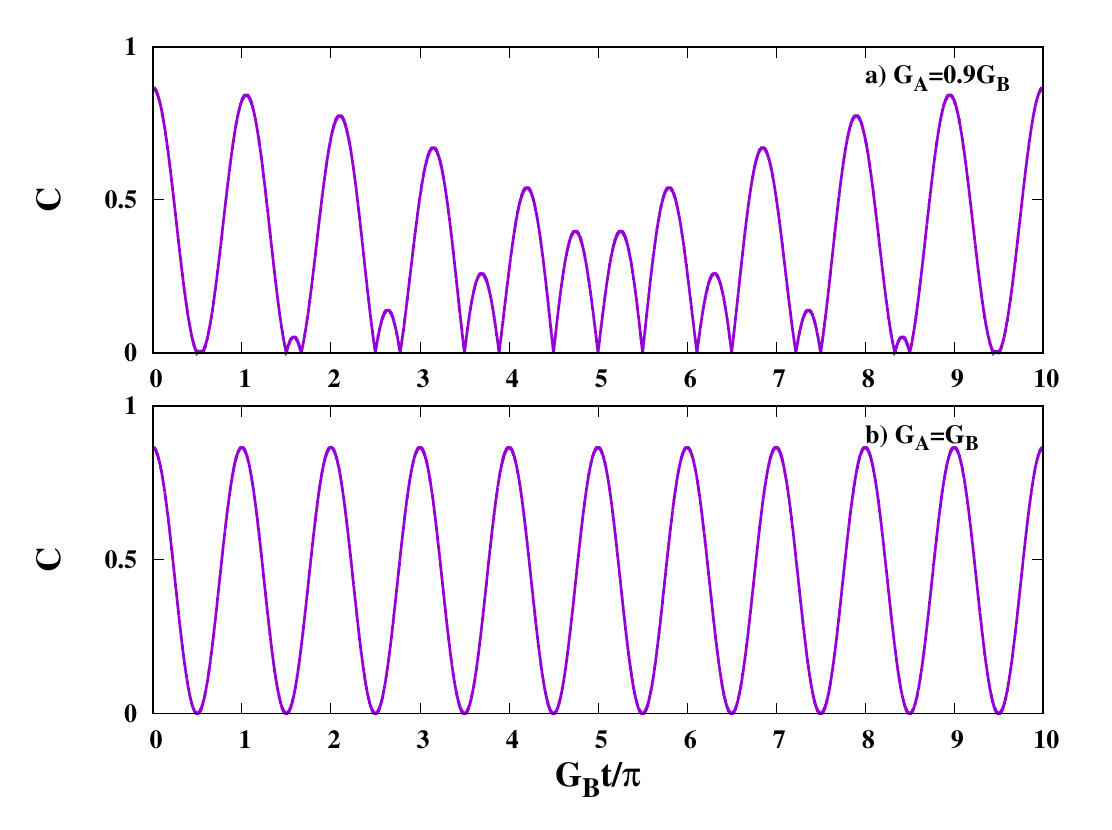} 
\caption{Atom-atom entanglement in the time evolved state of the double Jaynes-Cummings model. We exhibit here the behavior with respect to time of the atom-atom entanglement, as quantified by concurrence, within the double JC model, when the evolution starts off from the state in Eq. (\ref{si_0}), for $\alpha=\frac{\pi}{6}$. We choose $G_{A}=0.9G_{B}$ in diagram $(a)$, while $G_{A}=G_{B}$ in diagram $(b)$. The vertical axes represent concurrence and are measured in ebits, while the horizontal ones are of a dimensionless time.}
\label{con}
\end{figure} 
In {Fig. \ref{con}}, the time evolution of entanglement between the two atoms is exhibited for the cases when $G_{A}=G_{B}$ and also when $G_{A}$ is slightly different from $G_{B}$. 
\subsubsection{Response to disordered couplings}
Moving to the disordered case, we will now investigate the behavior of concurrence between the two atoms
in the presence of quenched disorder in coupling strengths. The Hamiltonian of the double JC model with disordered couplings is 
\begin{eqnarray}
\nonumber
&& \tilde{\mathcal{H}}=\frac{\hbar\omega}{2} \sigma_{z}^{A}+\hbar(1+\delta_A)(G_{A}\sigma_{+}^{A}a+G_{A}^{*}\sigma_{-}^{A}a^{\dagger})+\hbar \nu a^{\dagger}a \\ 
&& + \frac{\hbar\omega}{2} \sigma_{z}^{B} +\hbar(1+\delta_B) (G_{B}\sigma_{+}^{B}b + G_{B}^{*}\sigma_{-}^{B}b^{\dagger}) + \hbar\nu b^{\dagger}b,
\end{eqnarray}
where $\delta_A$ and $\delta_B$ are quenched disordered system parameters. For a given realization of $\delta_A$ and $\delta_B$, the atom-atom concurrence is 
\begin{equation}
C^{\delta_A \delta_B}(t)=|\sin 2 \alpha\cos(G_A(1+\delta_{A})t)\cos(G_B(1+\delta_{B})t)|.
\label{bristi}
\end{equation}
We will now  choose $\delta_{A}$ and $\delta_{B}$ from different types of distributions.\par
\noindent $\bullet$ \textbf{Gaussian quenched disorder:} In this case, we choose $\delta_{A}$ and $\delta_{B}$ randomly and independently from Gaussian distributions.
\begin{figure}
\includegraphics[width=9cm,height=6.5cm]{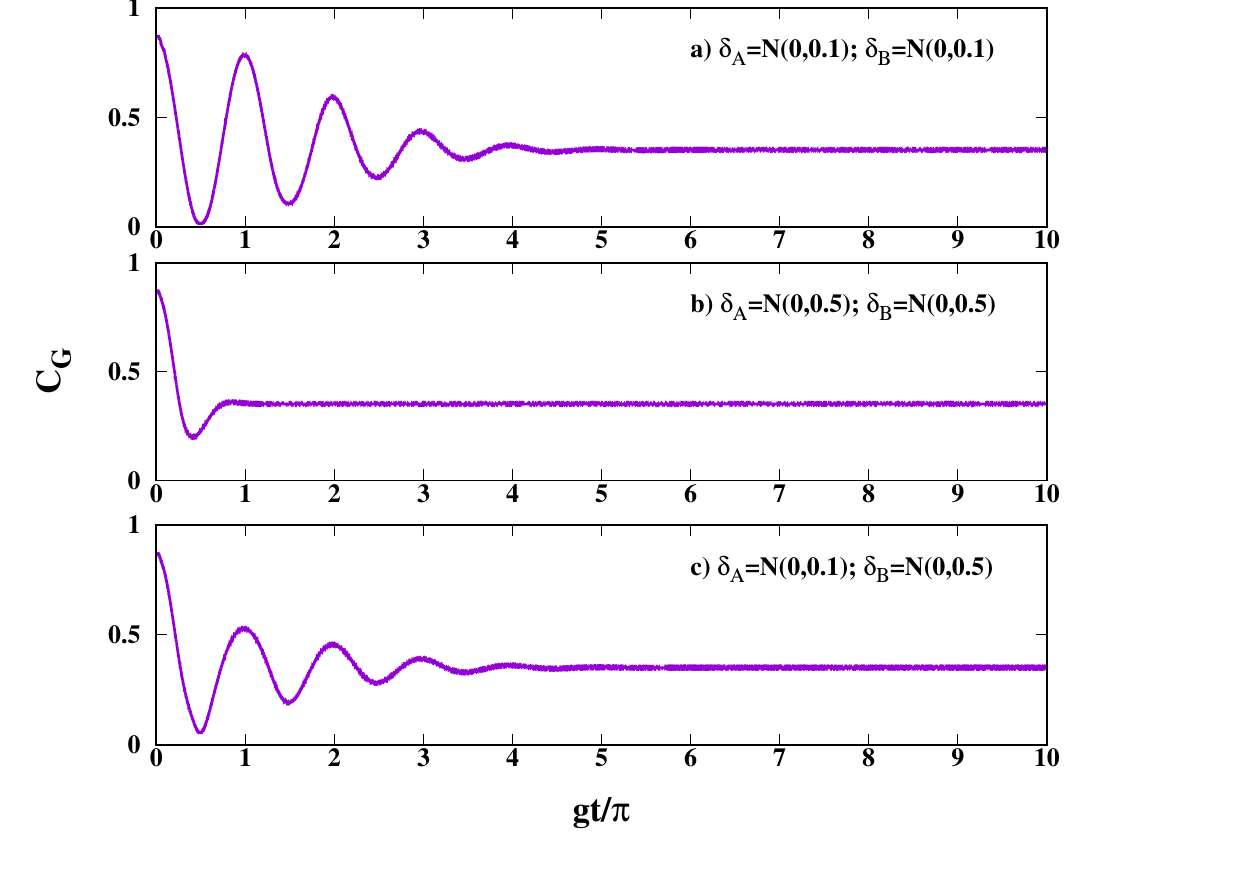} 
\caption{Atom-atom entanglement in the double JC model quickly attains a nonzero steady-state value in response to Gaussian quenched disorder. This is for the initial state corresponding to which the clean case is devoid of entanglement sudden death. 
The constant steady-state entanglement in the disordered case is to be compared with the 
strongly oscillating behavior of entanglement in time in the ordered case. See Fig. \ref{con}.
This can be seen as an advantage of the insertion of disorder for practical utilization of 
the atom-atom entanglement. In the ordered case, to obtain a high entanglement, we would be required to 
``freeze'' the system at certain specific times. However, in the disordered case, the freezing mechanism is automatically provided by the system dynamics, as a steady-state entanglement, not varying in time for moderately high times, is present, although its value is about half the maximal entanglement reachable in the ordered case. 
The vertical axes of the diagrams represent quenched averaged concurrence, measured in ebits, while the horizontal axes represent a dimensionless time. $\delta_{A}$ and $\delta_{B}$ are independently Gaussian disordered with mean zero and certain (nonzero) standard deviations. The different diagrams are for different sets of standard deviations, as marked on them. The notation $\delta_{A}=N(0,0.1)$ implies that $\delta_{A}$ is chosen randomly from a Gaussian (i.e., normal) distribution with mean zero and standard deviation $0.1$. The situation is the same for $(\delta_B)$. Here $G_A=G_B=g$.}
\label{con_gau} 
\includegraphics[width=9cm,height=6.5cm]{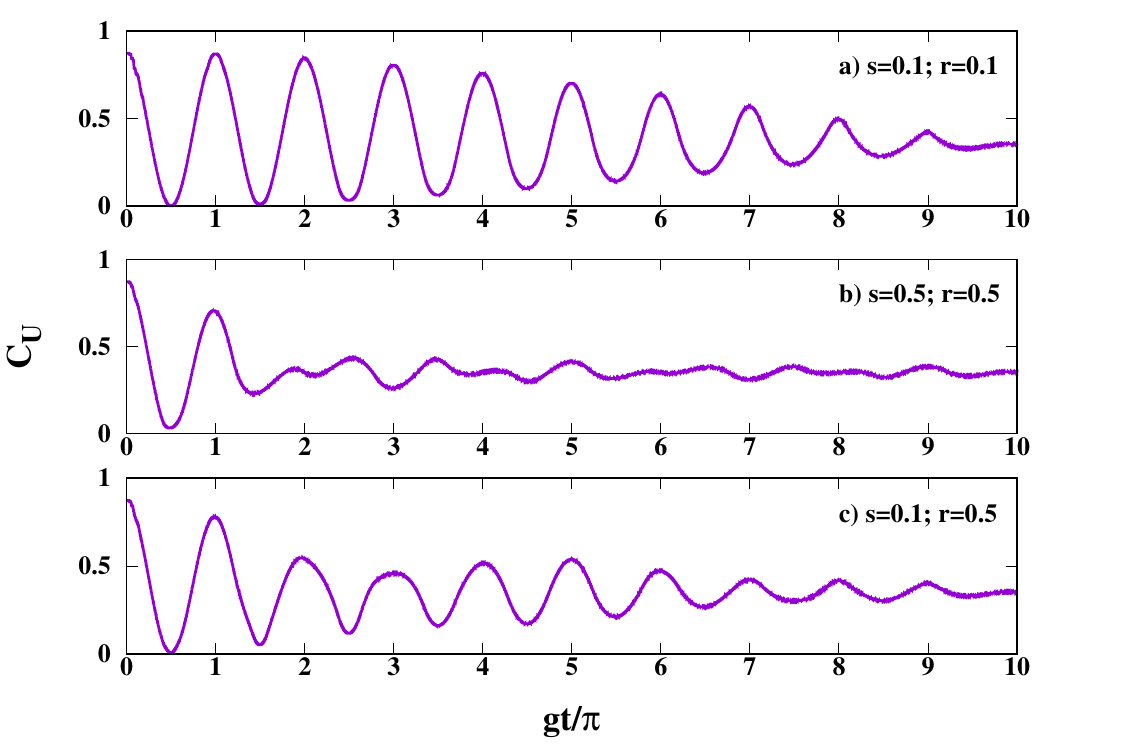} 
\caption{Response of atom-atom entanglement to uniform quenched disorder. The considerations are the same as in Fig. \ref{con_gau}, except that the disorders are uniform, and the values of $r$ and $s$ mentioned in each diagram refer to the standard deviations of $\delta_{A}$ and $ \delta_{B}$ used for the plot in that diagram.}
\label{con_uni}
\end{figure}
%
In {Fig. \ref{con_gau}}, we have shown the nature of the time evolution of quenched disordered entanglement for three cases. In general, the short-time quenched averaged entanglement has oscillations, which decrease in time to reach a steady value. The latter is nearly equal to the average concurrence in the corresponding case without disorder, although in the clean case the oscillations do not die out with time. How fast the average value is reached depends on the standard deviations of $\delta_{A}$ and $ \delta_{B} $. The larger the standard deviations, the faster the concurrence reaches the average value. This nature is quite similar to that in the single JC model with Gaussian disordered atom-cavity coupling, in which the population inversion and also the atom-photon entanglement approached a steady-state value, but the speed of the approach depended on the standard deviation of the disordered coupling strength.\par
\noindent $\bullet$ \textbf{Uniform quenched disorder:} Here we choose $\delta_{A}$ and $ \delta_{B}$ randomly and independently from uniform distributions in the range $[-\frac{s}{2}$ , $\frac{s}{2}]$ and $[-\frac{r}{2}$, $\frac{r}{2}]$, respectively.
\begin{figure}
\includegraphics[width=9cm,height=7cm]{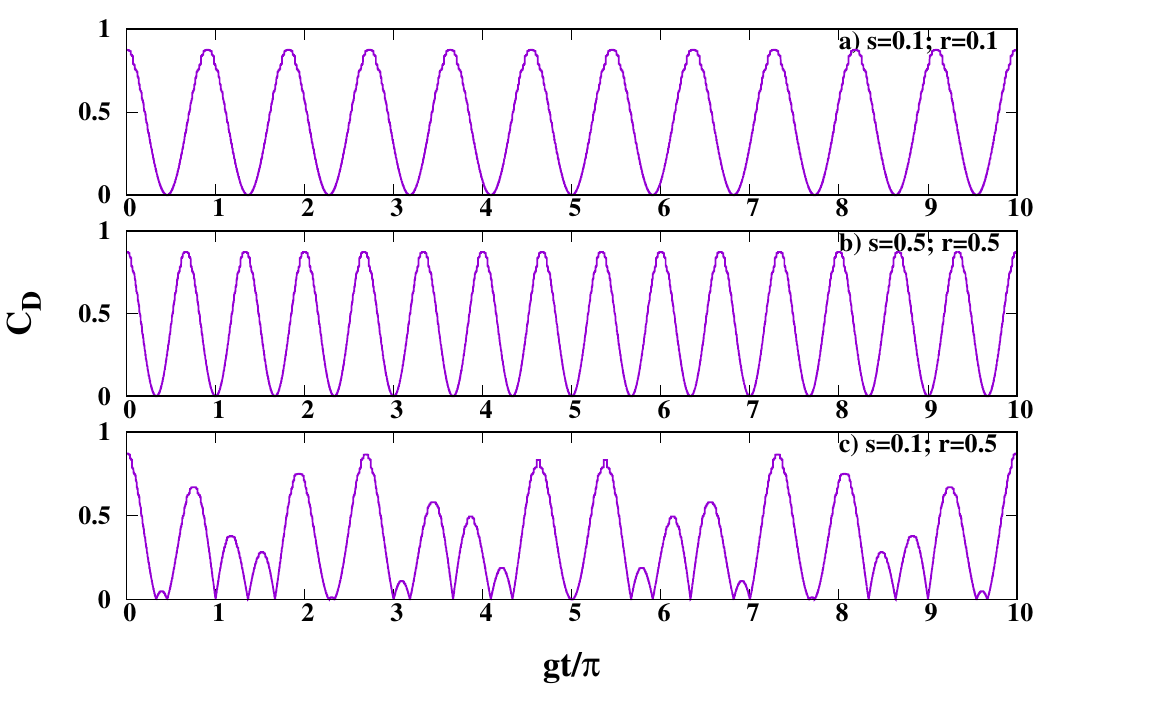} 
\caption{Time evolution of atom-atom concurrence with discrete quenched disordered couplings. The considerations are the same as in Fig. \ref{con_gau}, except that here the disordered couplings are independent discrete random variables. Precisely, $\delta_{A}$ and $\delta_{B}$ are independent discrete random variables taking up values $\pm\frac{s}{2}$ and $\pm\frac{r}{2}$, respectively, and with equal probabilities. The different values of $s$ and $r$ used in the different diagrams are marked therein.}
\label{con_dis}
\end{figure}
%
%
In {Fig. \ref{con_uni}}, we can see that introduction of a uniform disorder also shows a decay to a steady value for the time evolved quenched averaged entanglement, but the rate of decay is weaker than that for Gaussian quenched disorder. The rate depends on the values of the disorder strengths. Once again, the larger the disorder strength (as measured by the corresponding standard deviation), the quicker is the suppression of amplitude of oscillations. Like for Gaussian disorders, the steady-state value is again similar to the average of the clean case.\par
\noindent $\bullet$ \textbf{Discrete quenched disorder:} Here, $\delta_{A}$ and $\delta_{B}$ are chosen from the two-element sets $\lbrace -\frac{s}{2},\frac{s}{2}\rbrace$ and $\lbrace -\frac{r}{2},\frac{r}{2}\rbrace$, respectively, with equal probabilities of having $ -\frac{s}{2} (-\frac{r}{2})$ and $ \frac{s}{2} (\frac{r}{2})$.

{Fig. \ref{con_dis}} shows the behavior of quenched averaged concurrence with discrete quenched disorder. It does not show any sign of decay with time in response to the insertion of the quenched disordered couplings. The frequency of the oscillations increases with increase in standard deviations of the disordered couplings. If the values of the disorder strengths for the two $\delta$'s are different, then we get revivals with nonuniform amplitudes.\par
\begin{figure}
\includegraphics[width=9cm,height=7cm]{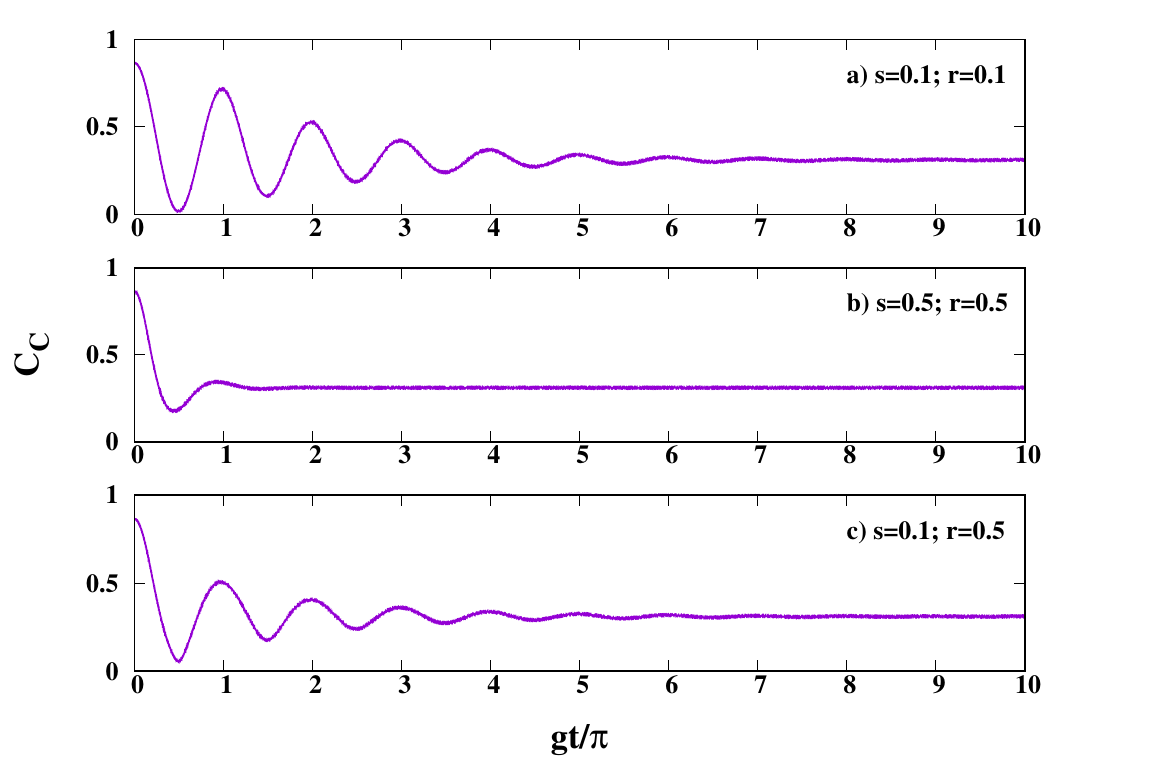} 
\caption{Decay of atom-atom entanglement oscillations in response to Cauchy-Lorentz disordered couplings. The quenched averaged concurrence is calculated by using the median of the concurrence distribution generated by the quenched disordered couplings. The semi-interquartile ranges of the Cauchy-Lorentz distributions of $\delta_{A}$ and $\delta_{B}$ are denoted by $s$ and $r$, respectively. All other considerations remain the same as in Fig. \ref{con_gau}. The profiles in the different diagrams in the figure are similar to the ones in the case of Gaussian disorder (see the corresponding diagrams in Fig. \ref{con_gau}), although the Cauchy-Lorentz disorder leads to slightly more oscillations before reaching the steady-state values.}
\label{con_cauchy} 
\end{figure}

\noindent $\bullet$ \textbf{Cauchy-Lorentz quenched disorder:} Here we have chosen $\delta_{A}$ and $\delta_{B}$ from Cauchy-Lorentz distributions and the quenched average is calculated by using the median of the corresponding distribution of atom-atom entanglement. In {Fig. \ref{con_cauchy}} we can see that the effect of Cauchy-Lorentz quenched disorder is almost the same as that of Gaussian quenched disorder. Compare with Fig. \ref{con_gau}. 

\subsection{When sudden death of entanglement is present in clean Hamiltonian}
\label{sub_2}
\subsubsection{Review of the clean case}
We now move over to the scenario where the initial state is so chosen that the clean Hamiltonian manifests the phenomenon of entanglement sudden death. For this case we will take a partially entangled atomic pure state which is a member of the span of the two Bell states $|\varphi_{\pm}\rangle$, and is written as
\begin{equation*}
|\varphi_{atom}\rangle=\cos \alpha |1_{A},1_{B}\rangle+\sin \alpha |0_{A},0_{B}\rangle, 
\end{equation*}
where $|\varphi_{\pm}\rangle=\frac{1}{\sqrt{2}}(|00\rangle+|11\rangle).$
So, the initial state for the total system is
\begin{equation}
|\varphi_{0}\rangle=(\cos \alpha |1_{A},1_{B}\rangle+\sin \alpha |0_{A},0_{B}\rangle)\otimes |0_a0_b\rangle.
\label{A}
\end{equation}
Now, using Eq. (\ref{2}) and Eq. (\ref{3}), the wave function at time $t$ is
\begin{eqnarray}
\nonumber
|\varphi_{t}\rangle=&&[\cos\alpha e^{-i\omega t} \lbrace \cos(G_{A}t)|1_{A},0_{a}\rangle -i \sin(G_{A}t)|0_{A},1_{a}\rangle \rbrace\\\nonumber
&&\otimes \lbrace \cos(G_{B}t)|1_{B},0_{b}\rangle-i \sin(G_{B}t)|0_{B},1_{b}\rangle \rbrace\\
&&+\sin \alpha e^{i\omega t} |0_{A},0_{a}\rangle \otimes |0_{B},0_{b}\rangle].
\end{eqnarray}
Tracing out the cavity parts, we have the density matrix for the two-atom system as
\begin{equation}
  \rho_{AB}=
  \left[ {\begin{array}{cccc}
   e & 0 & 0 & h^{*} \\
   0 & f & 0 & 0 \\
   0 & 0 & g & 0 \\
   h & 0 & 0 & 1-e-f-g \\
  \end{array} } \right],
\end{equation}
where
\begin{eqnarray*}
e&=&\cos^{2}\alpha|\cos(G_{A}t)\cos(G_{B}t)|^{2}\\
f&=&\cos^{2}\alpha|\cos(G_{A}t)\sin(G_{B}t)|^{2}\\
g&=&\cos^{2}\alpha|\sin(G_{A}t)\cos(G_{B}t)|^{2}\\
f&=&\cos^{2}\alpha|\sin(G_{A}t)\sin(G_{B}t)|^{2}+\sin^{2}\alpha\\
h&=&\cos\alpha \sin\alpha \cos (G_At)\cos(G_Bt).
\end{eqnarray*} 
The atom-atom concurrence is therefore
\begin{equation}
\begin{aligned}
\tilde{C}(t)={}&\max\big\{0,|\sin 2 \alpha\cos(G_{A}t)\cos(G_{B}t)|\\
&-\frac{1}{2}\cos^{2} \alpha|\sin(2G_At)\sin(2G_Bt)|\big\}.
\end{aligned}
\end{equation}
\begin{figure}
\includegraphics[width=9cm,height=7cm]{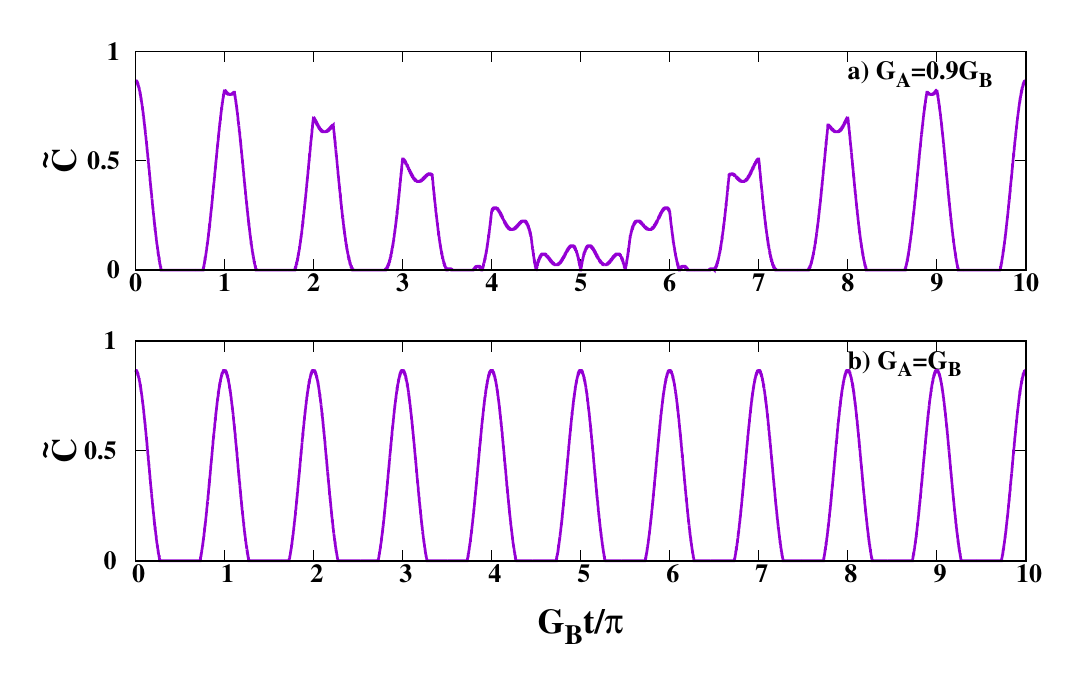}
\caption{Entanglement sudden death. See text for references where this phenomenon and the picture above were discussed, analyzed, and plotted. All considerations are the same as in Fig. \ref{con}, although the profiles are significantly different. Especially, there are semiperiodically appearing points on the time axis where entanglement vanishes with a discontinuous derivative, remains zero for a finite span of time, and becomes nonzero (``revival") again, and again with a discontinuous derivative. Also, the initial state of evolution is a different one here. See text.}
\label{con_1}
\end{figure}
In {Fig. \ref{con_1}}, we can see that there are sudden deaths of entanglement, followed by revivals, for both the cases when $G_{A}=G_{B}$ \cite{Vieira, Eberly, Chen, Pandit, Zhang, Shen} and when $G_{A}$ is slightly different from $G_{B}$ \cite{Shen, Sainz, Hu}. Here, by the phrase ``sudden death", it is meant that entanglement vanishes at a point in time in such a way that its derivative is discontinuous there.
Note that we can express \(\tilde{C}(t)\) as 
\[\max\{0, C(t) -\frac{1}{2}\cos^{2} \alpha|\sin(2G_At)\sin(2G_Bt)|\},\] where \(C(t)\) is the entanglement in the case when there is no sudden death, given by Eq. (\ref{banya}). From the expression for \(C(t)\), it is clear that it cannot exhibit entanglement sudden death. The positive quantity \(\frac{1}{2}\cos^{2} \alpha|\sin(2G_At)\sin(2G_Bt)|\) getting subtracted from \(C(t)\) creates the possibility that \(\tilde{C}(t)\) may exhibit entanglement sudden death, although this is not guaranteed, and depends on the relative values of \(C(t)\) and the 
additional quantity \(\frac{1}{2}\cos^{2} \alpha|\sin(2G_At)\sin(2G_Bt)|\).
\subsubsection{Response to disordered couplings}
As for the case where entanglement sudden death is absent, we now consider quenched disordered coupling strengths within the Hamiltonian $\tilde{\mathcal{H}}$. The concurrence, for a given realization of the disordered  parameters, is given by   
\begin{equation}
\begin{aligned}
&\tilde{C}^{\delta_A \delta_B}(t)=\\&{}\max\big\{0,|\sin 2 \alpha\cos(G_At(1+\delta_A))\cos(G_Bt(1+\delta_B))|\\
&-\frac{1}{2}\cos^{2} \alpha|\sin(2G_At(1+\delta_A))\sin(2G_Bt(1+\delta_B))|\big\}.
\end{aligned}
\end{equation}
Just like in the ordered case, we find that 
\(\tilde{C}^{\delta_A \delta_B}(t)\) can be expressed as 
\begin{equation}
\begin{aligned}
&\max\big\{0, C^{\delta_A \delta_B}(t)\\
&-\frac{1}{2}\cos^{2} \alpha|\sin(2G_At(1+\delta_A))\sin(2G_Bt(1+\delta_B))|\big\},
\end{aligned}
\end{equation}
where \(C^{\delta_A \delta_B}(t)\) is the entanglement for a particular configuration of disorder in the case when there is no sudden death, as given by Eq. (\ref{bristi}). An average over the disorder for
\(C^{\delta_A \delta_B}(t)\) cannot exhibit entanglement sudden death. However, the positive 
quantity, \(\frac{1}{2}\cos^{2} \alpha|\sin(2G_At(1+\delta_A))\sin(2G_Bt(1+\delta_B))|\), 
after averaging over the disorder will remain positive, and when subtracted from a disorder-averaged 
\(C^{\delta_A \delta_B}(t)\) creates the possibility of entanglement sudden death, even after disorder averaging. However, whether entanglement sudden death will actually be exhibited will depend on the relative values of the disorder-averaged \(C^{\delta_A \delta_B}(t)\) and \(\frac{1}{2}\cos^{2} \alpha|\sin(2G_At(1+\delta_A))\sin(2G_Bt(1+\delta_B))|\).\par
\noindent $\bullet$ \textbf{Gaussian quenched disorder:} The quenched averaged time evolved concurrence, corresponding to which the clean case has entanglement sudden death, has a behavior that is quite similar to the case for which the clean case does not exhibit the sudden death. An important difference is obtained when the disorder strengths are weak, and in such cases, for short times, the sudden death of the clean case persists in the disordered one.
\begin{figure}
\includegraphics[width=9cm,height=7cm]{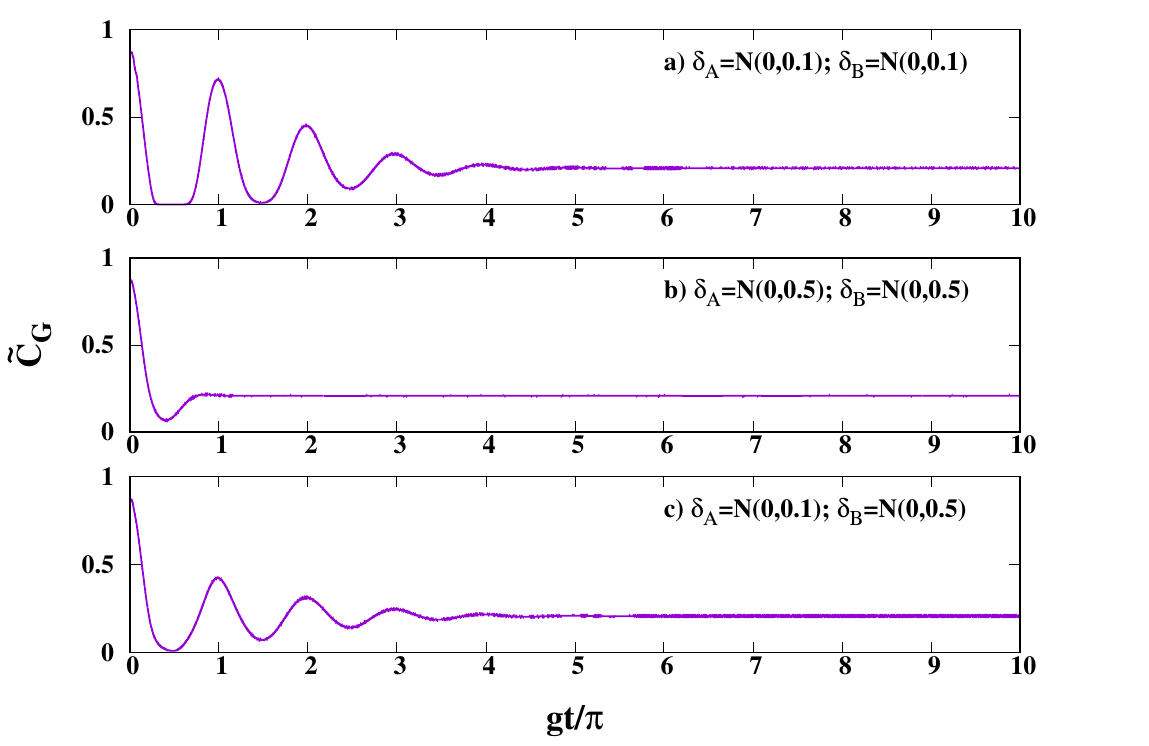}  
\caption{Response of entanglement sudden death to Gaussian quenched disorder in the double JC model. All considerations are the same as in Fig. \ref{con_gau}, except that here the initial state of evolution is given by Eq. (\ref{A}). The sudden death can be 
avoided,
unless the disorder strengths are very low, viz. standard deviations$\approx 0.1$ of $\delta_{A}$ and $\delta_{B}$. It is to be noted here that quenched averaging can transform an occurrence of entanglement sudden death into a noninflexion double root of the function given by the disorder-averaged entanglement with respect to time. Such instances in the above diagrams are to within numerical precision in our calculations.}   
\label{plot}
\end{figure} 
See Fig. \ref{plot}, and compare with Fig. \ref{con_gau}. Note that disorder averaging can lead 
to situations where the sudden death in the ordered case is transformed into a noninflexion double root of the profile of disorder-averaged entanglement as a function of time.\par
\noindent $\bullet$ \textbf{Uniform quenched disorder:}
\begin{figure}
\includegraphics[width=9cm,height=7cm]{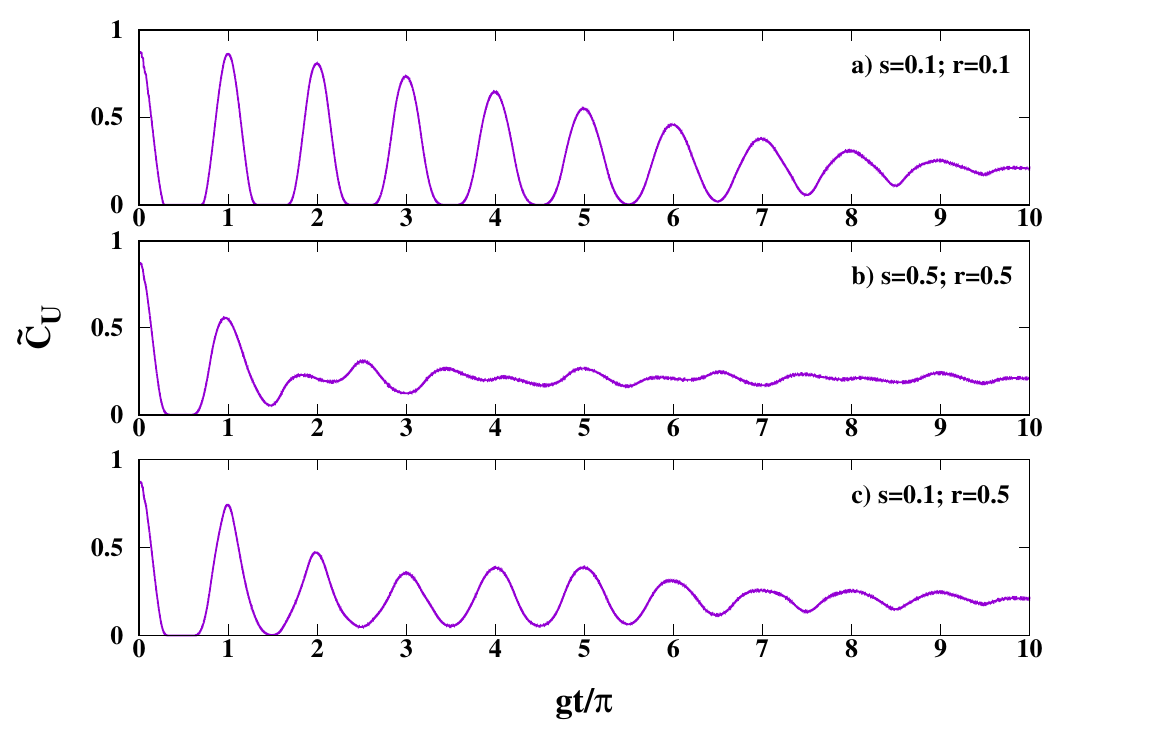} 
\caption{Higher disorder strength or longer time needed for uniform disorder than for Gaussian to wipe out entanglement sudden death. All considerations remain the same as in Fig. \ref{plot}, except that the disorders are uniform. The meanings of $r$ and $s$ in the different diagrams are as in Fig. \ref{con_uni}.}
\label{plot1}
\end{figure}
Fig. \ref{plot1} shows the nature of atom-atom concurrence in the presence of uniform quenched disorder in the coupling strengths. Just like for the Gaussian disorder, the general behavior, after quenched averaging, seems to obliterate the differences between the cases of the presence and absence of entanglement sudden death in the corresponding clean cases. The exceptions are when the disorder strengths are low or the time of observation is longer. However, the obliteration by using uniform quenched disorder requires more disorder strength than the Gaussian one. We remember that the standard deviation of a disorder is being used to quantify the strength of that disorder.\par
\begin{figure}
\includegraphics[width=9cm,height=7cm]{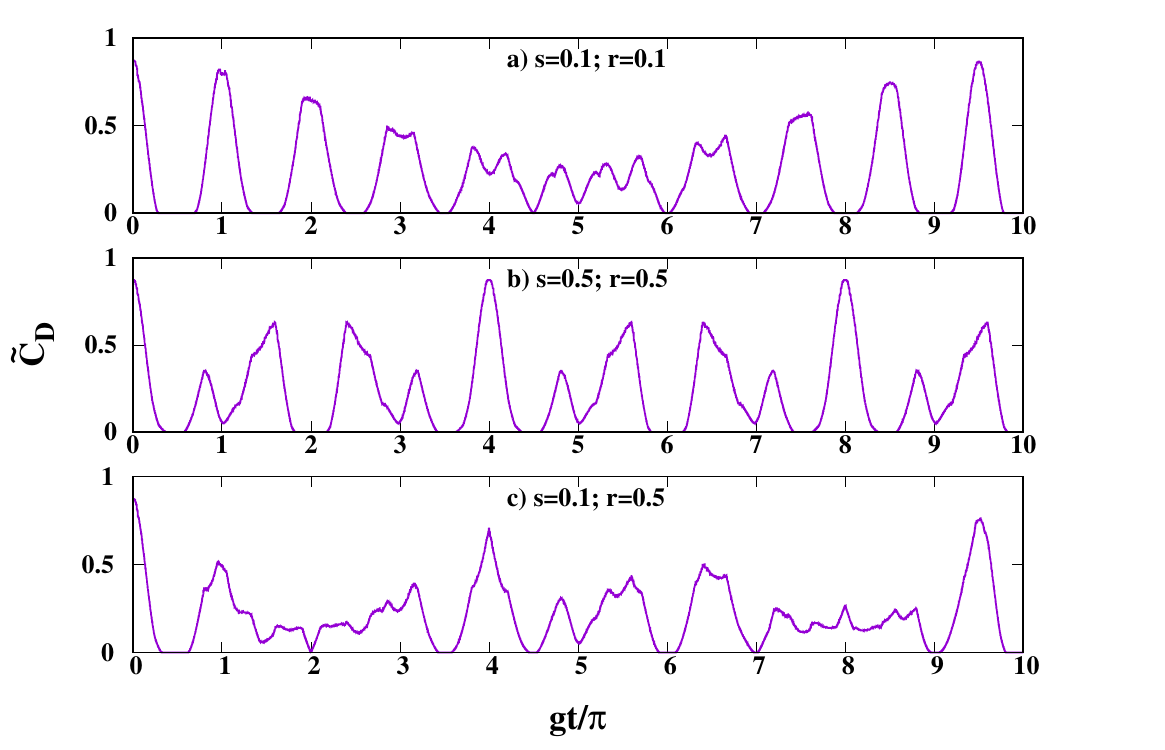} 
\caption{Persistence of entanglement sudden death despite onset of discrete quenched disorder. All considerations are the same as in Fig. \ref{plot}, except that the disorders are discrete. The meanings of $r$ and $s$ in the different diagrams are the same as in Fig. \ref{con_dis}.}
\label{plot2}
\end{figure}
\noindent $\bullet$ \textbf{Discrete quenched disorder:} Fig. \ref{plot2} shows the behavior of atom-atom concurrence in the presence of discrete quenched disorder. In this case, in contrast to the two previous cases of continuous disorders, the phenomenon of entanglement sudden death persists for much longer time spans and for much higher disorder strengths.\par
\noindent $\bullet$ \textbf{Cauchy-Lorentz quenched disorder:} Like in the previous cases, here also Cauchy-Lorentz quenched disorder garners a response that is similar to that for Gaussian disorder, although the rate of decrease of amplitude of oscillations is less than that for Gaussian disorder and also the phenomenon of entanglement sudden death persists to higher strengths of the disorder. Note that the strength of the Cauchy-Lorentz disorder is measured by using the semi-interquartile range, while that for the Gaussian one is quantified by employing the standard deviation. The quenched averaged entanglement is calculated by using the median.
\begin{figure}
\includegraphics[width=9cm,height=7cm]{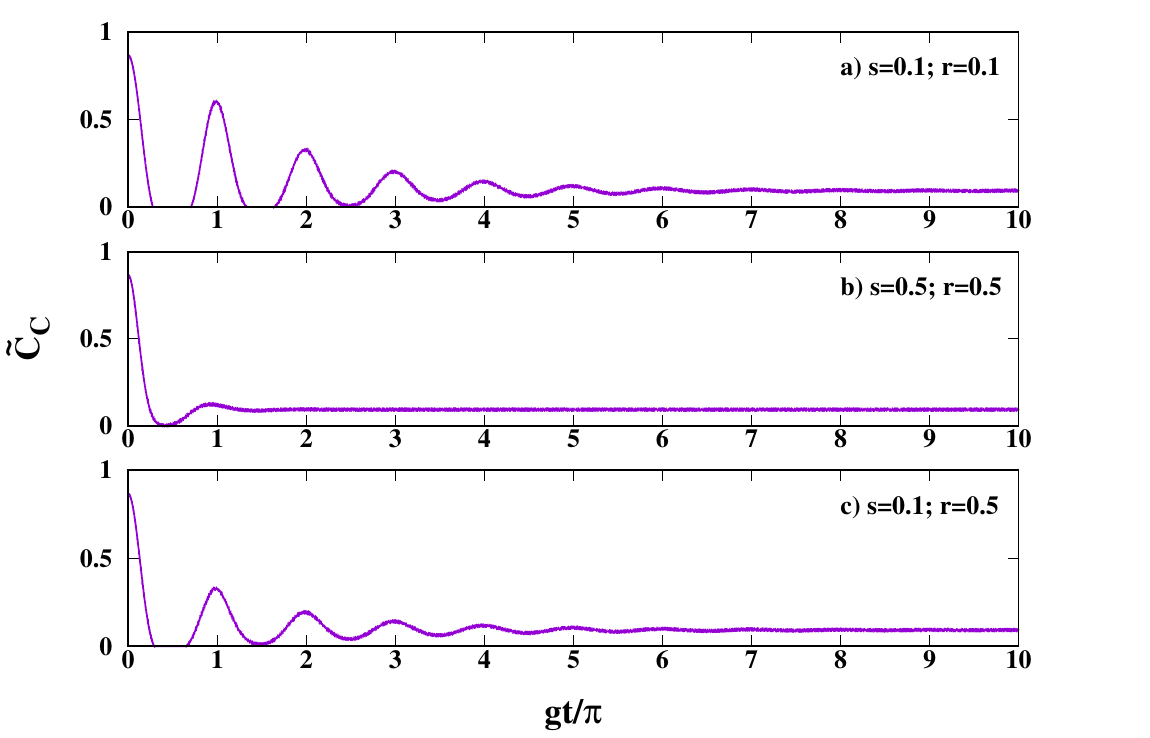}
\caption{Response of atom-atom entanglement to Cauchy-Lorentz disordered couplings in the double JC model. All considerations are the same as in Fig. \ref{plot}, except that the disorders are Cauchy-Lorentz. The meanings of $s$ and $r$ are as in Fig. \ref{con_cauchy}.}
\label{con_cauchy1}
\end{figure}
See Fig. \ref{con_cauchy1} for depictions in a few cases.\par
\begin{figure}
\includegraphics[width=9cm,height=6cm]{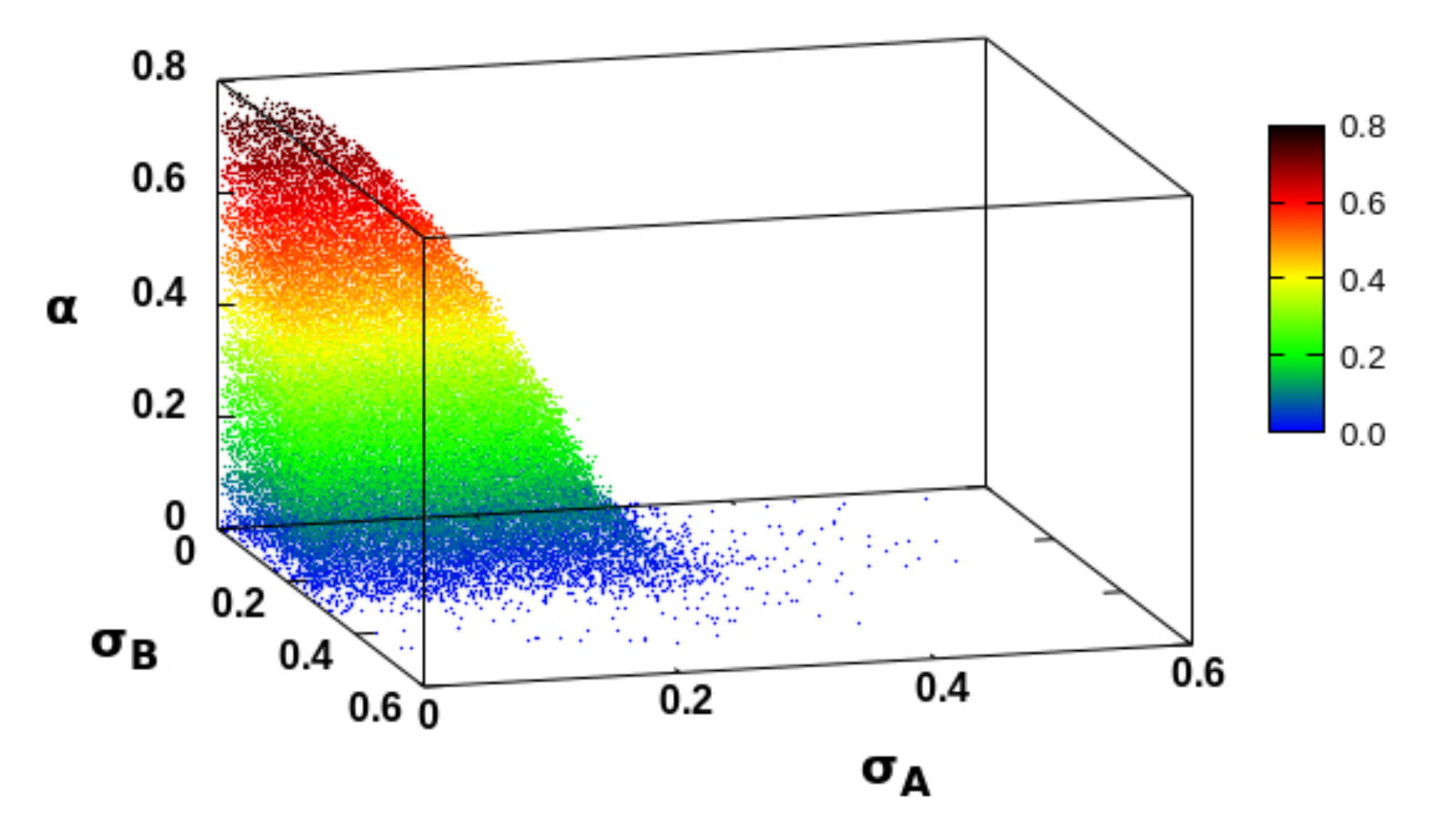}
\caption{How much disorder is needed to wipe out entanglement sudden death, and its relation to initial entanglement in the double JC model. The initial state of the evolution is given by Eq. (\ref{A}). For points in the marked region, the quenched averaged entanglement retains entanglement sudden death, while the same is absent in the remaining region. $\sigma_A$ and $\sigma_B$ are standard deviations of the Gaussian-distributed $\delta_A$ and $\delta_B$, the means of the latter being both zero. $\alpha$ quantifies the amount of entanglement in the initial atom-atom state. All quantities are dimensionless.}
\label{reg_gau}  
\end{figure}
\begin{figure}
\includegraphics[width=9cm,height=6cm]{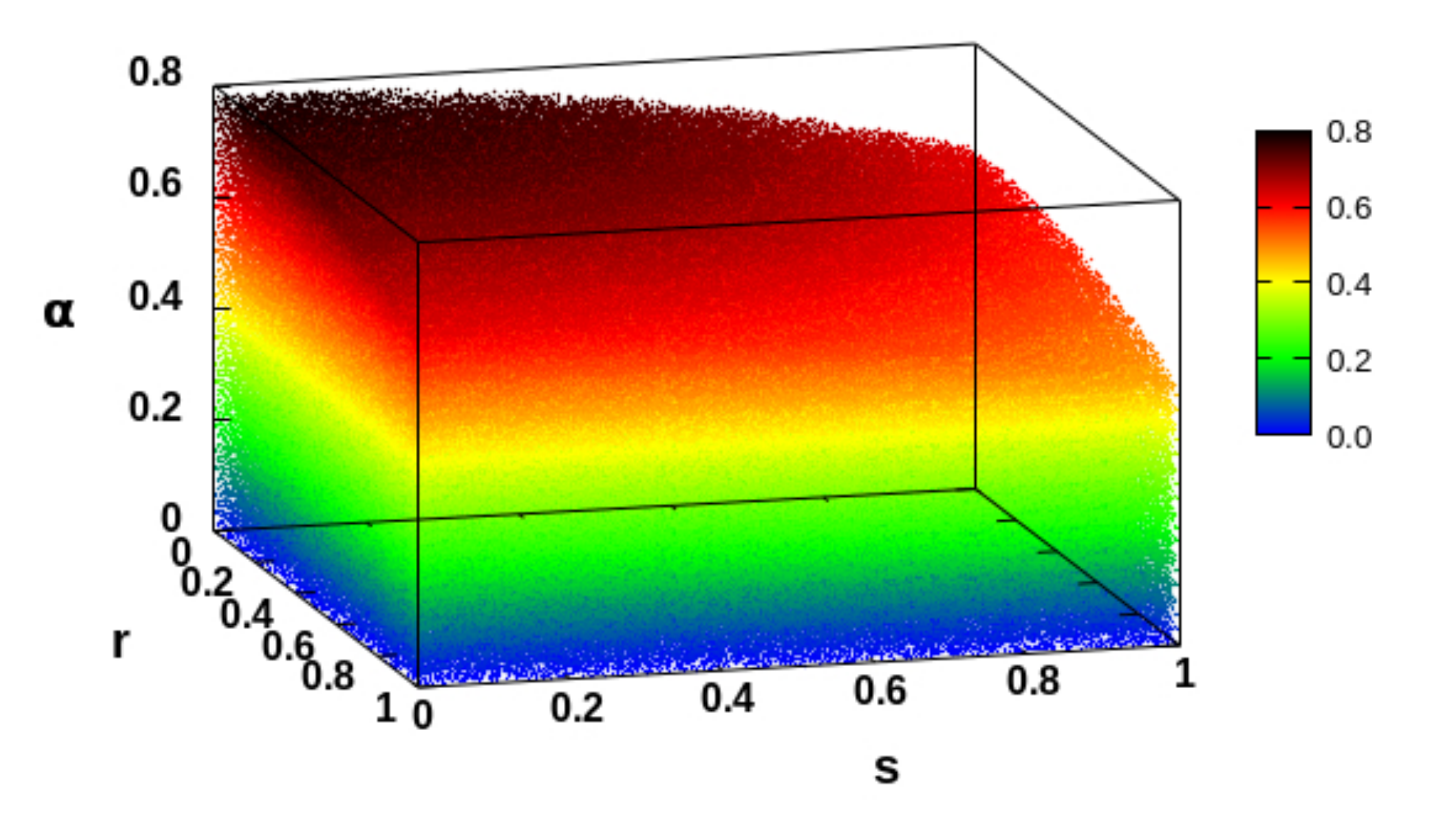}
\caption{Absence and presence of entanglement sudden death in quenched averaged entanglement for uniform disordered couplings in the double JC model. The considerations are the same as in Fig. \ref{reg_gau}, except that the disorders are uniform, with $s$ and $r$ being standard deviations of $\delta_A$ and $\delta_B$, respectively. The analysis is carried out for $0\leq s, r\leq 1$.}
\label{reg_uni}  
\end{figure}
\begin{figure}
\includegraphics[width=9cm,height=6cm]{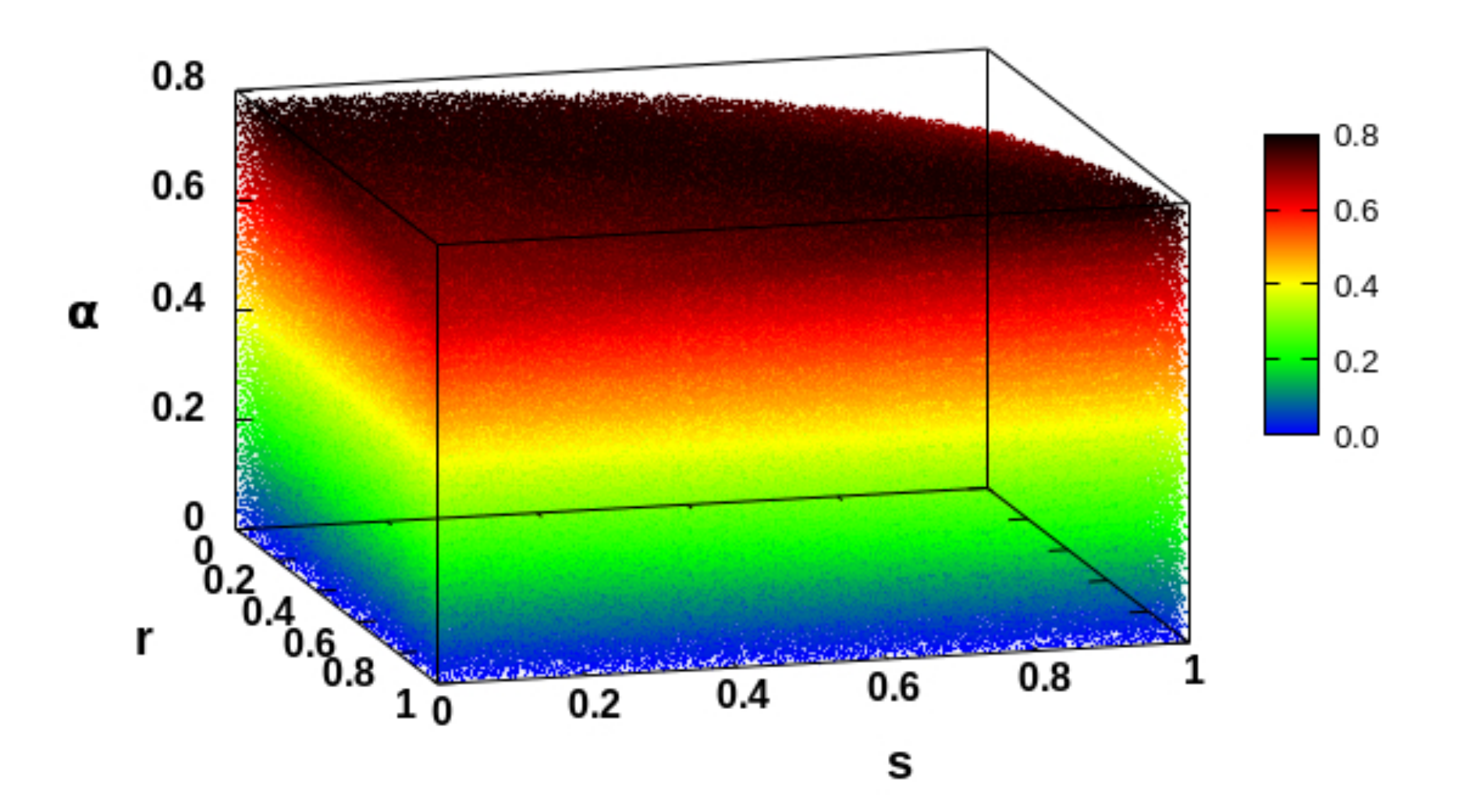}
\caption{Disorder strengths vs initial entanglement for entanglement sudden death against discrete quenched disordered couplings in the double JC model. The considerations are the same as in Fig. \ref{reg_gau}, except that the disorders are discrete, with $s$ and $r$ being standard deviations of $\delta_A$ and $\delta_B$, respectively. We have considered values of $s$ and $r$ in the range [0,1].}
\label{reg_dis}  
\end{figure}
\begin{figure}
\includegraphics[width=9cm,height=6cm]{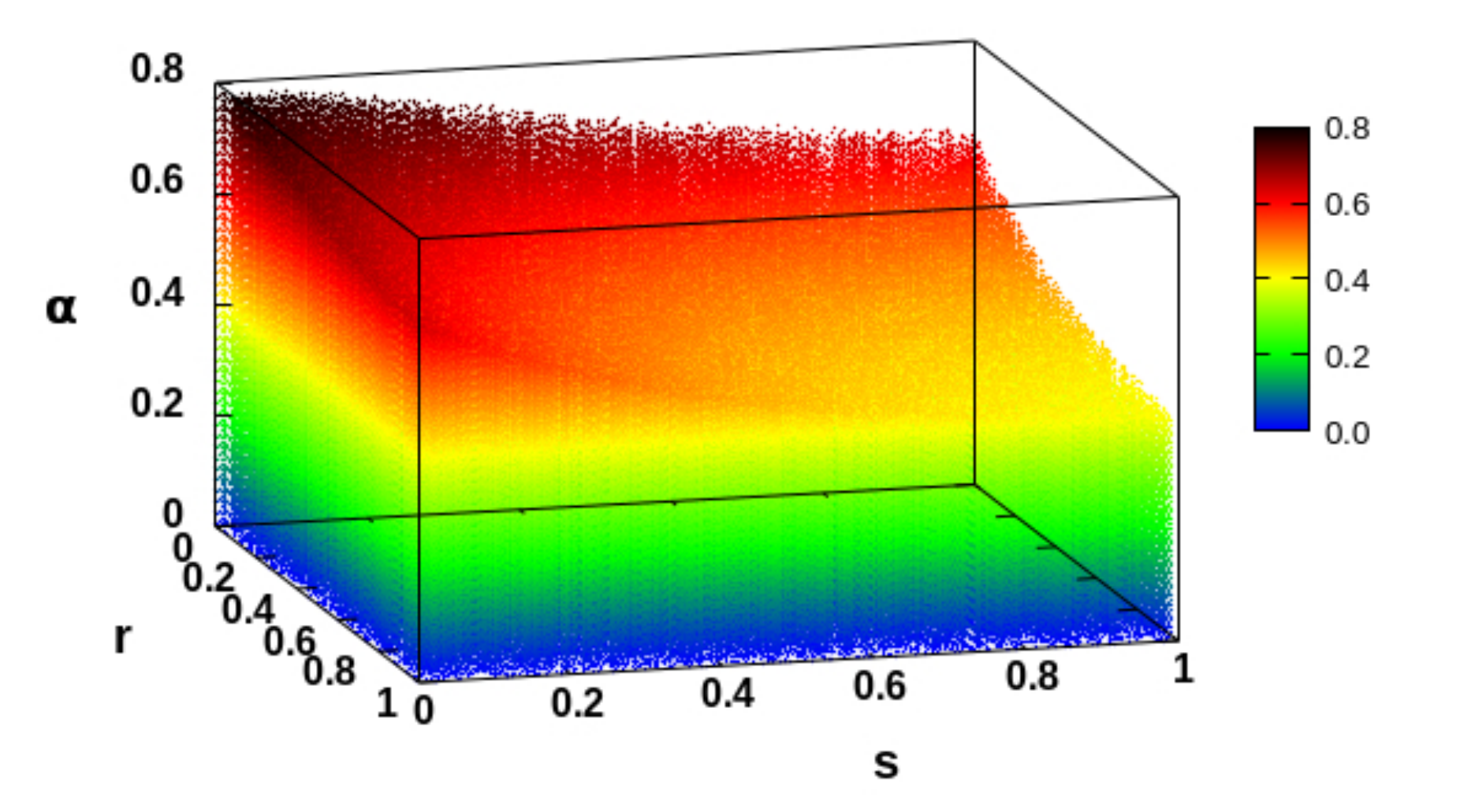}
\caption{Median-based quenched averaged entanglement for Cauchy-Lorentz disordered couplings in the double JC model with respect to entanglement sudden death. The considerations are the same as in Fig. \ref{reg_gau}, except that the disorders are Cauchy-Lorentz, with $s$ and $r$ being semi-interquartile ranges of $\delta_A$ and $\delta_B$, respectively. Also, the quenched averaging is performed by considering the median instead of the mean. The semi-interquartile ranges are taken in [0,1].}
\label{reg_cau} 
\end{figure}
\textbf{Region of disorder and atomic parameters sustaining entanglement sudden death:} 
Let us now identify the regions in the space of disorder and atomic parameters that support entanglement sudden death, even after quenched averaging. By ``disorder parameters", we mean the strengths of the disorders inserted in the atom-cavity couplings. By ``atomic parameter", we mean the parameter $\alpha$ of the initial state [in Eq. (\ref{A})] of the two atoms. In this three-dimensional space, we find out the region for which entanglement sudden death persists, vis-\`a-vis the region which does not support the same. These regions, for the four different types of disorders considered in this paper, are depicted in Figs. \ref{reg_gau}, \ref{reg_uni}, \ref{reg_dis} and \ref{reg_cau}. Except for small differences, the regions are quite similar for uniform, discrete, and Cauchy-Lorentz disorders. The region for Gaussian disorder is, however, significantly different and much smaller. The following note is in order here. In all the previous discussions, we have seen that Cauchy-Lorentz disorder inflicts a similar effect, qualitatively, on the system characteristics as Gaussian disorder. There were, however, quantitative differences. A measure of that difference is seen in the difference in volumes of the regions where entanglement sudden death is present, after quenched averaging, in Figs. \ref{reg_gau} and \ref{reg_cau}.\\ 
\section{Entanglement in time evolution of quenched disordered Double Jaynes-Cummings model with atom-atom coupling term present in the Hamiltonian}
\label{atom-atom}
In this section, we want to see the effect of atom-atom coupling terms on the phenomenon of entanglement sudden death in the double Jaynes-Cummings model. We have investigated the response of entanglement sudden death to the separate introduction of two prototypical atom-atom interaction terms, viz., $1)$ the Ising interaction term and $2)$ the anisotropic $XY$ spin-exchange interaction term.
\subsection{In presence of Ising interaction}
\begin{figure}
\includegraphics[width=9cm,height=7cm]{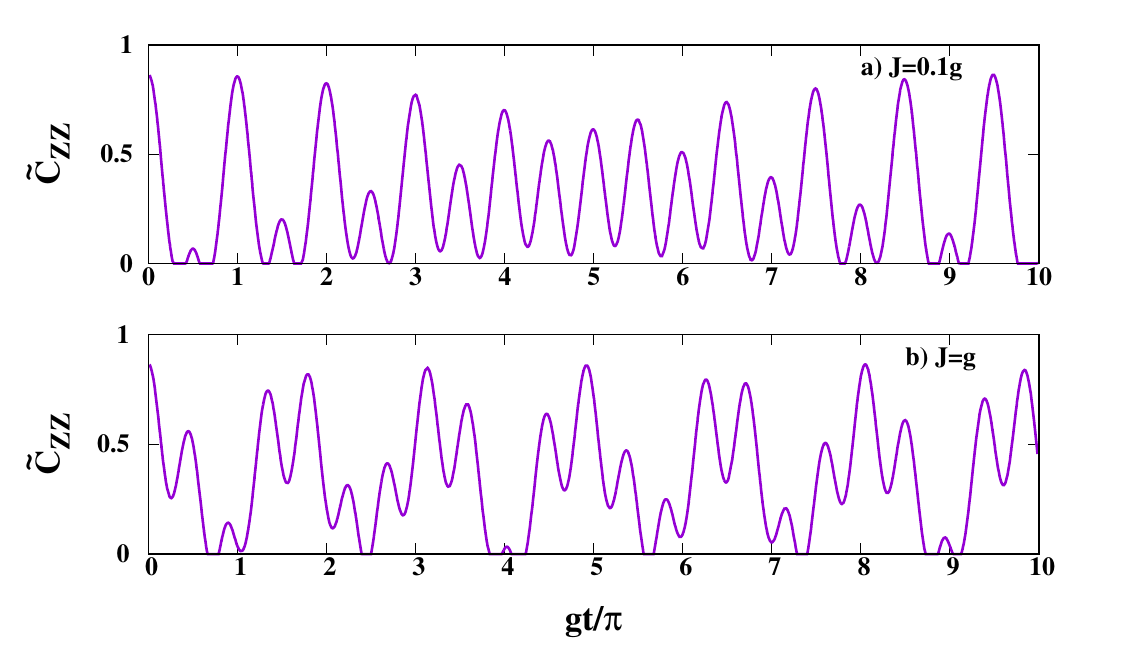}
\caption{Entanglement sudden death in the presence of the atom-atom coupling term. All considerations are the same as in Fig. \ref{con}. Only, in this case, $G_A=G_B=g$ in both the diagrams, and $J_z\neq 0$. We have chosen $J_z=0.1g$ in diagram $a)$ and $J_z=g$ in diagram $b)$. The profiles are significantly different from the sudden death in Fig \ref{con_1}$b)$, that corresponded to $J_z=0$. The sudden death of entanglement is delayed here, and also split, depending on the value of $J_z$. See text for description.}
\label{ZZ_WD}
\end{figure}

If an Ising interaction term is present, then the modified Hamiltonian in the clean case will be
\begin{equation}
\mathcal{H}^{\prime}=\mathcal{H}+J_{z}\hbar(\sigma_{z}^{A}\otimes\sigma_{z}^{B}).
\end{equation}
where $\mathcal{H}$ is given by Eq. (\ref{H}). We choose $G_A=G_B=g$. $J_z$ is proportional to the Ising coupling strength between the two atoms. The initial state for the case where sudden death of entanglement is present is represented by $|\varphi_0\rangle$, given in Eq. (\ref{A}). After an evolution governed by $\mathcal{H}^{\prime}$ we get the evolved state as
\begin{eqnarray}
\nonumber
|\varphi_{t}^{\prime}\rangle&=&a_1|1_A1_B0_a0_b\rangle+a_2|0_A0_B1_a1_b\rangle+a_3|1_A0_B0_a1_b\rangle\\
&+&a_4|1_A0_B0_a1_b\rangle+a_5|0_A0_B0_a0_b\rangle
\end{eqnarray}
where the coefficients $a_1$, $a_2$, $a_3$, $a_4$, and $a_5$ are evaluated for two cases: $i)\:J_z=0.1g$ and $ii)\:J_z=g$. The atom-atom entanglements for these two cases are depicted in Fig. \ref{ZZ_WD}, where we have again considered concurrence as a measure of entanglement. The corresponding plot for $J_z=0$ is presented in Fig. \ref{con_1}. $b)$. We find that a small $J_z$, viz., $J_z=0.1g$, results in a splitting of the first region of ESD for $J_z=0$. For later times, the introduction of $J_z$ can even remove an ESD of $J_z=0$. For larger $J_z$, even the onset of ESD can be delayed in time with respect to the $J_z=0$ case. In both the cases, the profiles are semiperiodic.\par
\begin{figure}
\includegraphics[width=9cm,height=6.5cm]{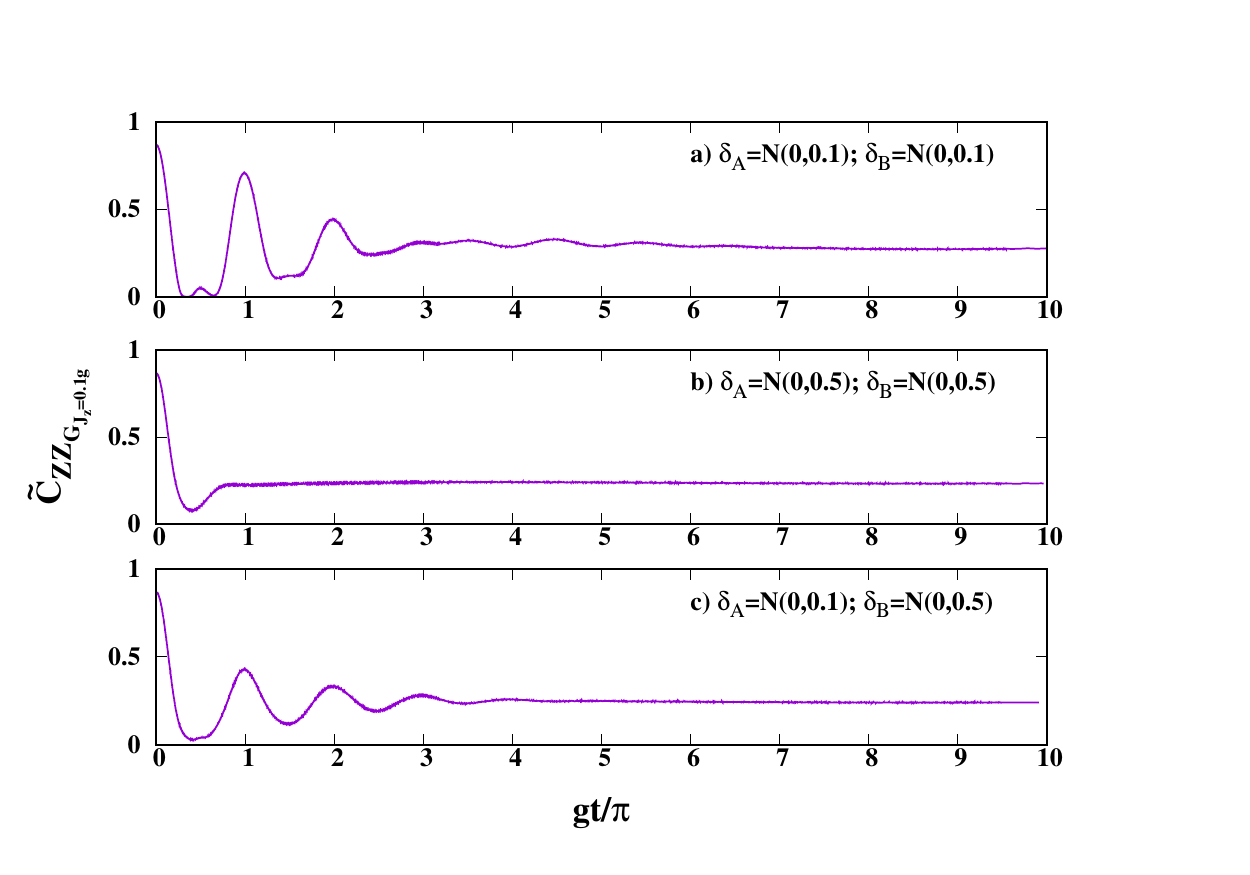}
\caption{Entanglement sudden death of atom-atom entanglement of the double Jaynes-Cummings model in the presence of the Ising interaction term for $J_z=0.1g$. All considerations are the same as in Fig. \ref{plot}, expect that $J_z\neq 0$ here. Compare with Figs. \ref{plot}, \ref{ZZ_WD}$a)$, and \ref{ZZ_G_J=g}. See text for a description on the response of ESD to the insertion of the $J_z$ coupling. In particular, the saturation value of entanglement is approximately between $0.23$ and $0.27$, while in Fig. \ref{plot} the saturation value of entanglement is $\approx 0.207$.}
\label{ZZ_G_J=0.1g}
\includegraphics[width=9cm,height=6.5cm]{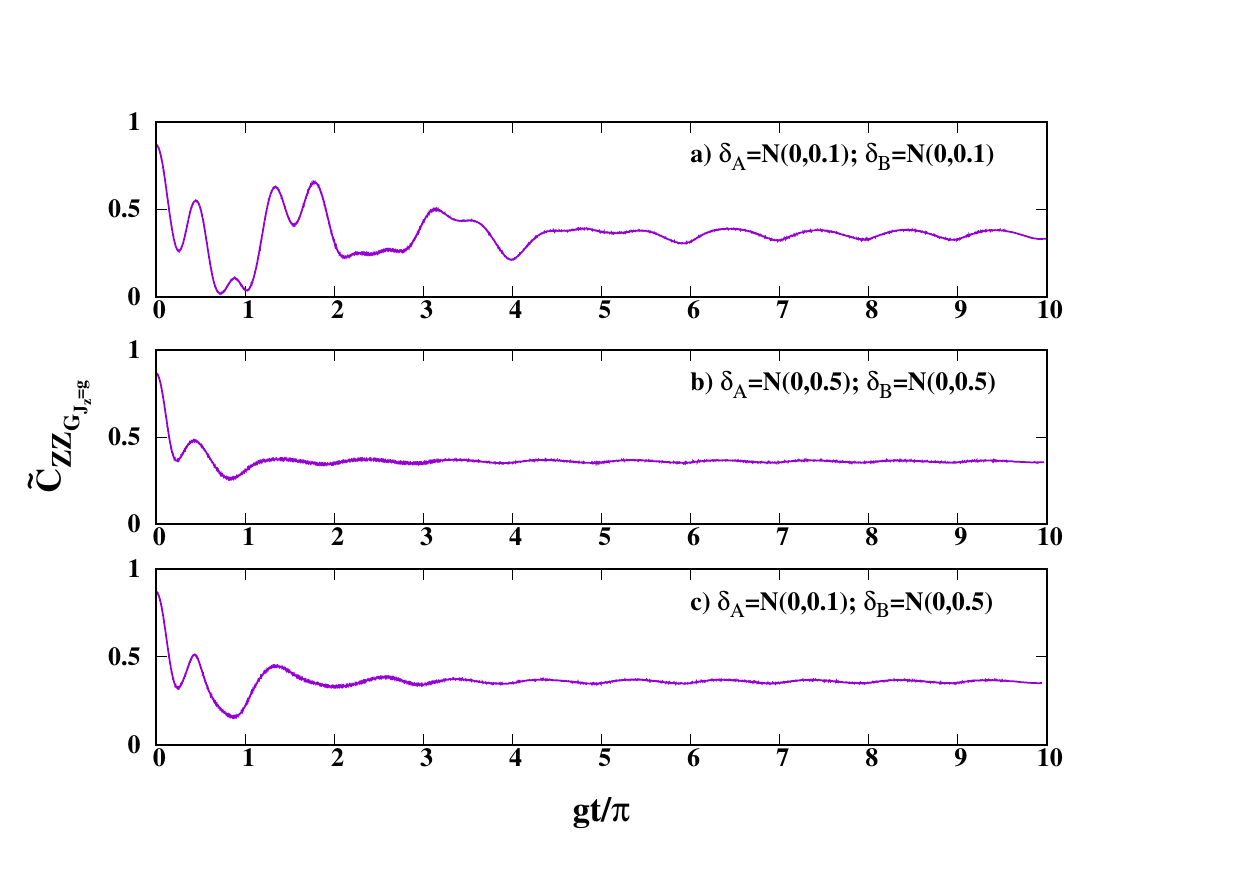}
\caption{Entanglement sudden death of atom-atom entanglement of the double Jaynes-Cummings model in the presence of the Ising interaction term for $J_z=g$. All considerations are the same as in Fig. \ref{plot}, except for the value of $J_z$. In the first diagram atom-atom entanglement is not yet saturated for the time span considered although the oscillations are dying out, and the saturation value of entanglement is $\approx 0.35$ for diagrams $b)$ and $c)$ .}
\label{ZZ_G_J=g}
\end{figure}

We now investigate the behavior of sudden death of entanglement with atom-atom coupling in the presence of Gaussian quenched disorder in the atom-photon couplings. As before, we consider quenched disordered atom-photon coupling strengths so that in the clean Hamiltonian $G_A$ is replaced by $g(1+\delta_A)$ and $G_B$ is replaced by $g(1+\delta_B)$. $\delta_A$ and $\delta_B$ are chosen independently from Gaussian distributions for a large numbers of realizations, from which we get the quenched averaged entanglement as a function of time. In Figs. \ref{ZZ_G_J=0.1g} and \ref{ZZ_G_J=g}, the evolution of atom-atom entanglement in the presence of the Ising interaction term is exhibited for two values of $J_z$. We can see that for the larger value of $J_z$ sudden death of entanglement is wiped out for smaller values of disorder strength than in the low $J_z$ case. However the saturation value of entanglement is attained at longer times for larger $J_z$.
\subsection{In presence of anisotropic XY spin-exchange interaction}
\begin{figure}
\includegraphics[width=9cm,height=4cm]{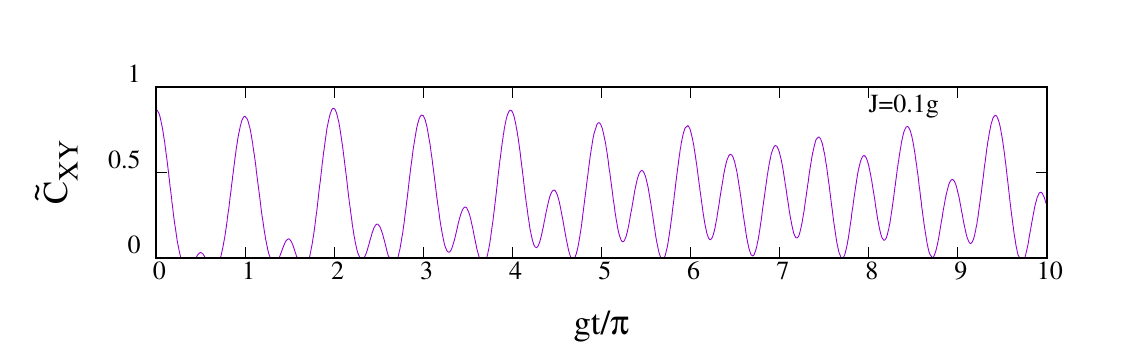}
\caption{Entanglement sudden death in the double Jaynes-Cummings model in the presence of anisotropic $XY$ spin-exchange interactions. We choose $\gamma=0.5$. All considerations are the same as in Fig. \ref{ZZ_WD}, except that there is an $XY$ interaction term between the atoms.}
\label{XY_WD}
\end{figure}

We next consider the case where interaction between the two atoms is the spin-exchange interaction governed by the anisotropic $XY$ Hamiltonian. The double Jaynes-Cummings Hamiltonian with this interaction term is
\begin{equation}
\mathcal{H}''=\mathcal{H}+\hbar(J_{x}\sigma_{x}^{A}\otimes\sigma_{x}^{B}+J_{y}\sigma_{y}^{A}\otimes\sigma_{y}^{B}),
\end{equation}
where $\mathcal{H}$ is given by Eq. (\ref{H}). $J_x$ and $J_y$ are the coupling constants in $x$ and $y$ directions, and $J_x=J(1+\gamma)$, $J_y=J(1-\gamma)$, where $\gamma$ is a constant. In this case we have made an assumption to attain numerical tractability, as the effective basis of the Hamiltonian has an infinite number of elements. As the system is closed, the total energy of the system is conserved. Here we have taken a maximum of two excitations in the cavity states. This assumption reduces the elements of the effective basis of the Hamiltonian. The elements of this basis are $|1_A1_B0_a0_b\rangle$, $|0_A1_B1_a0_b\rangle$, $|1_A0_B0_a1_b\rangle$, $|1_A0_B1_a0_b\rangle$, $|0_A1_B0_a1_b\rangle$, $|0_A0_B1_a1_b\rangle$, $|0_A0_B0_a2_b\rangle$, $|0_A0_B2_a0_b\rangle$, and $|0_A0_B0_a0_b\rangle$. The initial state of the evolution is taken to be $|\varphi_0\rangle$, as given by Eq. (\ref{A}), and then the evolved state at time $t$ is given by
\begin{equation}
\begin{aligned}
\label{eq:37}
|\varphi''_t\rangle&=b_1|1_A1_B0_a0_b\rangle+b_2|0_A1_B1_a0_b\rangle+b_3|1_A0_B0_a1_b\rangle\\
&+b_4|1_A0_B1_a0_b\rangle+b_5|0_A1_B0_a1_b\rangle+b_6|0_A0_B1_a1_b\rangle\\
&+b_7|0_A0_B0_a2_b\rangle+b_8|0_A0_B2_a0_b\rangle+b_9|0_A0_B0_a0_b\rangle.
\end{aligned} 
\end{equation}
It is the introduction of the spin-exchange interaction term in the Hamiltonian that leads to the possibility of higher excitations in the cavities. We have, however, truncated the actual Hilbert space into an effective Hilbert space which has at most two excitations in the cavity modes. It is plausible that this assumption is only valid when the relative strength of the spin-exchange interaction is small, i.e., when \(J/g\) is small. As an example, we analyze the case when \(J/g = 0.1\). The nature of atom-atom entanglement in the absence of disorder is exhibited in Fig. \ref{XY_WD}.
\begin{figure}
\includegraphics[width=9cm,height=7cm]{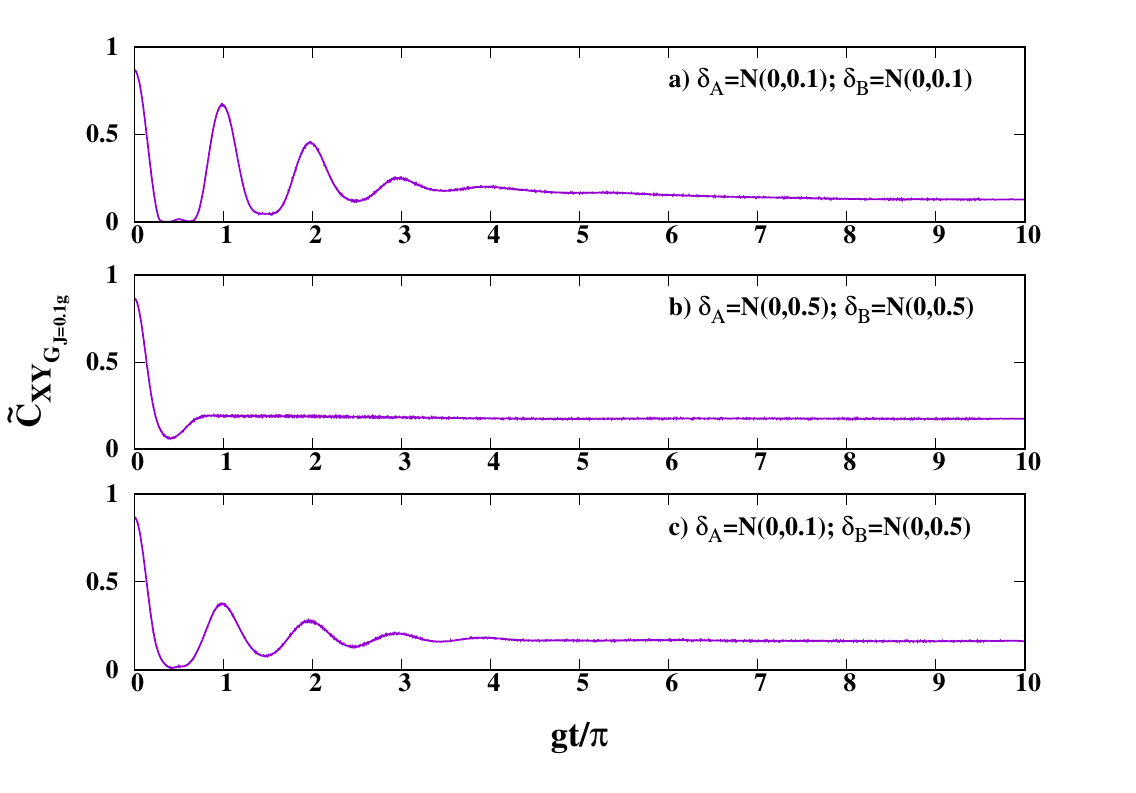}
\caption{Atom-atom entanglement in the presence of the anisotropic $XY$ spin-exchange interaction term. All considerations are the same as in Fig. \ref{ZZ_G_J=0.1g}, except that the atom-atom interaction is now an $XY$ one with $J=0.1g$ and $\gamma=0.5$. As in all previous cases, the vanishing of sudden death of entanglement depends on the strengths of the disorder parameters. The large-time saturation value of entanglement for diagram $a)$ is $\approx 0.13$, while for diagram $b)$ and $c)$ it is $\approx 0.17$.}
\label{XY_J=0.1g_G}
\end{figure}
We will now consider quenched disordered atom-photon couplings in the presence of the small spin-exchange interaction. The disorders are applied for the atom-photon couplings, and are the same as for the Hamiltonian \(\mathcal{H}\) of Eq. (34) in the preceding subsection. The disorders are again Gaussian distributed. The quenched averaged atom-atom entanglement, as quantified by concurrence, is plotted in the diagrams in Fig. \ref{XY_J=0.1g_G}, for different strengths of the disorders. The behaviors are broadly similar to those obtained for the Ising interaction between the atoms. Compare with the diagrams in Fig. \ref{ZZ_G_J=0.1g}.         
\section{Conclusion}
\label{9}
We have been concerned in this paper with the effects of archetypal forms of quenched disorder in atom-cavity coupling constants on the population inversion and the entanglement of single and double Jaynes-Cummings models. We have considered Gaussian as well as non-Gaussian models of disorder, and the system characteristics were investigated in their quenched averaged versions. The non-Gaussian distributions examined are uniform, discrete, and Cauchy-Lorentz ones. The analysis for the Cauchy-Lorentz distributed disorder necessitated the consideration of median-based quenched averages, which we have introduced, before its examination.\par
We began with the Jaynes-Cummings model of a single two-level system and a single mode of an electromagnetic field, for which we analyzed the response to disorder in atom-cavity interaction of the population inversion as well as the atom-photon entanglement. We found that Gaussian disorder strongly suppresses the collapse and revival phenomenon of population inversion, while for non-Gaussian disorder, the suppression is milder. For the atom-photon entanglement, we found in particular that the same can have nontrivial oscillations even when the population inversion has been suppressed.\par
For the double Jaynes-Cummings model, we focused our attention on the atom-atom entanglement, and its response to quenched disorder in the atom-cavity interactions. There are certain classes of entangled initial
atom-atom states that lead to the phenomenon of entanglement sudden death in the clean double Jaynes-Cummings model. We looked at the effect of quenched disorder in atom-atom entanglement in the cases when the phenomenon is present as well as those in which the same is absent. In particular, we provided the minimal values of the disorder strengths that, for a given initial entanglement in the atom-atom quantum state, will wipe out the possibility of entanglement sudden death. We also investigate the response of atom-atom entanglement of a double Jaynes-Cummings model in the presence of an additional atom-atom coupling term, with the latter being modeled, separately, by Ising and $XY$ interactions.


We acknowledge useful comments received from Jonas Larson. We thank Anirban Pathak for pointing out a typo in Eq. (\ref{eq:37}) in a previous version.

\end{document}